\documentclass[aoas]{imsart}

\RequirePackage{amsthm,amsmath,amsfonts,amssymb}
\RequirePackage[authoryear]{natbib}%% uncomment this for author-year citations

\startlocaldefs
\usepackage{amsmath}
\usepackage{graphicx}
\usepackage{enumerate}
\usepackage{natbib}
\usepackage{url}
\usepackage{bm}
\usepackage{dsfont}
\usepackage{amssymb}
\usepackage{hhline}
\usepackage{multirow}
\usepackage{xcolor}
\usepackage{subfig}
\usepackage{xr}
\usepackage{adjustbox}
\usepackage{changepage}
\usepackage{lipsum}

\newcommand{\bbeta}{\mbox{\boldmath $\beta$}}
\renewcommand{\hat}{\widehat}

\newcommand{\btheta}{\bm{\theta}}
\newcommand{\E}{\mathds{E}}

\def\T{{\sf T}}

\let\hat\widehat

\DeclareMathOperator*{\ind}{1{\hskip -2.5 pt}\hbox{I}}  % Indicator
\newcommand{\R}{\mathds{R}}
\newcommand{\be}{\bm{e}}

\newcommand{\bx}{\bm{x}}
\newcommand{\by}{\bm{y}}
\newcommand{\bz}{\bm{z}}
\newcommand{\bX}{\bm{X}}

\newcommand{\bY}{\bm{Y}}
\newcommand{\bZ}{\bm{Z}}

\newcommand{\balpha}{\bm{\alpha}}
\newcommand{\bgamma}{\bm{\gamma}}
\newcommand{\bepsilon}{\bm{\epsilon}}

\newcommand{\bphi}{\bm{\phi}}

\newcommand{\bSigma}{\bm{\Sigma}}

\endlocaldefs

\begin{document}

\begin{frontmatter}
\title{Characterizing Alzheimer's Disease Biomarker Cascades Through Non-linear Mixed Effect Models}
%\title{A sample article title with some additional note\thanksref{T1}}
\runtitle{AD Biomarker Cascades}
%\thankstext{T1}{A sample of additional note to the title.}

\begin{aug}
%%%%%%%%%%%%%%%%%%%%%%%%%%%%%%%%%%%%%%%%%%%%%%%
%% Only one address is permitted per author. %%
%% Only division, organization and e-mail is %%
%% included in the address.                  %%
%% Additional information can be included in %%
\author[A]{\fnms{Zhuojun}~\snm{Tang}},
\author[B, C]{\fnms{Yuxin}~\snm{Zhu}\ead[label=e1]{Co-first author}},
\author[D]{\fnms{Kexin}~\snm{Zhang}}
\and
\author[E, C]{\fnms{Zheyu}~\snm{Wang}\ead[label=e4]{Correspondence: wangzy@jhu.edu}}
%%%%%%%%%%%%%%%%%%%%%%%%%%%%%%%%%%%%%%%%%%%%%%
%% Addresses                                %%
%%%%%%%%%%%%%%%%%%%%%%%%%%%%%%%%%%%%%%%%%%%%%%
\address[A]{Department of Mathematics, University of Pittsburgh}

\address[B]{Department of Neurology, Johns Hopkins University\printead[presep={,\ }]{e1}}

\address[C]{Department of Biostatistics, Johns Hopkins University}

\address[D]{Department of Applied Mathematics and Statistics, Johns Hopkins University}

\address[E]{Division of Quantitative Sciences, Sidney Kimmel Comprehensive Cancer Center, Johns Hopkins University\printead[presep={,\ }]{e4}}
\end{aug}

\begin{abstract}
Alzheimer's Disease (AD) research has shifted to focus on biomarker trajectories and their potential use in understanding the underlying AD-related pathological process. A conceptual framework was proposed in modern AD research that hypothesized biomarker cascades as a result of underlying AD pathology. In this paper, we leveraged the idea of biomarker cascades and used a non-linear mixed effect model to depict AD biomarker trajectories as a function of the latent AD disease progression. We tailored our methods to address a number of real-data challenges that are often present in AD studies. Simulation studies were performed to investigate the proposed approach under various realistic but less-than-ideal situations. Finally, we illustrated the methods using real data from the BIOCARD and the ADNI studies. The analyses investigated cascading patterns of AD biomarkers in these datasets and presented prediction results for individual-level profiles over time. These findings highlight the potential of the conceptual biomarker cascade framework to be leveraged for diagnosis and monitoring.
\end{abstract}
%non-linear mixed effects, latent variables, absence of a gold standard, disease modeling, biomarker monitoring
% \begin{keyword}[class=MSC]
% \kwd[Primary ]{}
% \kwd{}
% \kwd[; secondary ]{}
% \end{keyword}


\end{frontmatter}

\newpage
%\spacingset{1.9} % DON'T change the spacing!

\section{Introduction}\label{sec:introduction}
%Research on Alzheimer's disease (AD) has shifted towards focusing on the preclinical stage of AD, where the pathophysiological process has just begun and therapeutic interventions are most likely to be successful \citep{sperling2011toward}. 

%The preclinical stage of Alzheimer's disease can last for decades, during which patients experience deterioration in some biomarkers but may not exhibit any clinical signs of cognitive decline that warrant a clinical diagnosis. In 2011, the National Institute on Aging and the Alzheimer's Association proposed a research framework that centers around defining AD based on biomarkers collected from study participants to identify the underlying disease process \citep{sperling2011toward,mckhann2011diagnosis, albert2011diagnosis}. This new biomarker-based disease process framework suggests that underlying AD processes 1) should be modeled as a continuous process instead of a series of discrete changes and 2) emphasize the role of biomarkers through clinically observable measurements showing the progression of the underlying disease and through multifactorial profiles of people who are in preclinical, mild cognitive impairment, or dementia stages. According to the Jack model of hypothesized biomarker cascades \citep{jack2010hypothetical}, A$\beta$ biomarkers become abnormal first while individuals are still cognitively normal, followed by neurodegenerative biomarkers and cognitive symptoms, forming S-shaped progression cascades along the disease progression scale. 
Research on Alzheimer's Disease (AD) has shifted its focus on early stages of AD, during which cognitive impairment has yet started or is minimal, and therefore interventions have the greatest potential to be effective. Because patients appear to be normal with little clinical signs in these early stages, researchers are especially interested in understanding the measurable AD biomarkers in relation to the underlying AD pathological stages. To this end, Jack and colleagues proposed an influential framework that described a hypothetical AD biomarker progression pattern along stages of AD \citep{jack2010hypothetical, jack2013update}. This conceptual framework, commonly referred to as the \textit{Jack model}, assumes that: 1) the underlying AD pathological process is continuous and clinical decline happens gradually; 2) AD biomarkers become abnormal and reach a plateau in a temporal order, possibly following sigmoid-shaped trajectories, forming the so-called \textit{biomarker cascades}; and 3) the AD biomarker trajectories are important to be studied in relation to the underlying disease process and can potentially be used for effective identification of preclinical and mild cognitive impairment stages of AD. The Jack model provides a unique opportunity for developing data-driven models on AD biomarkers cascade to inform AD staging. In particular, AD studies with biomarker measurements collected when participants were cognitively normal without any sign of clinical symptoms are especially valuable because they might contain information on biomarker profiles during the preclinical stage of AD. Combining such early-stage AD biomarker data with longitudinal observations could potentially further reveal how AD biomarkers change during various AD stages.

The Jack model has inspired numerous statistical models and methods. Many of these models studied the biomarker trajectories as sigmoid functions of age or other forms of chronological time \citep{li2019bayesian, donohue2014estimating, sun2018nonlinear, jedynak2012computational, amieva2008prodromal}. To account for individual heterogeneity, some models incorporated random effects that allow for individual variations in the timing of biomarker deterioration, such as through individual time shifts of the biomarker trajectory \citep{li2019bayesian, donohue2014estimating}, or in inflection points which capture the period of the greatest rate of change \citep{sun2018nonlinear}. These methods essentially model disease progression as an effect of time on biomarker deterioration, in addition to the effect of healthy aging. As a result, one potential limitation of these approaches is that they lack a clearly defined disease progression beyond the levels of biomarker measurements. The significant heterogeneity observed in biomarker measurements among individuals in the same disease stages, as well as the variability in the pattern of biomarker deterioration, make it difficult to fully capture the potential fluctuations and dynamics of the AD pathophysiology process over time \citep{jack2013update}. Recognizing these challenges, Jack and his colleague used the phrase ``distance traveled along the pathophysiological pathway'' instead of a chronological time in their updated model \citep{jack2013update}. It motivated a series of work that utilized latent variables to model the underlying disease progress that varies with time and manifests in biomarker measurements.  Most of these latent variable models considered the disease as a discrete variable and assumed that biomarkers are influenced by this latent discrete disease \citep{proust2014joint, sun2018nonlinear, wang2018biomarker}. While these methods are may offer simplicity when linking with clinical decisions, they do not fully capture the continuous spectrum of disease progression from normality to abnormality. Other approaches exist that assume an underlying continuous latent variable, focusing on constructing an individual-level disease score \citep{jedynak2012computational} or understanding domain-specific latent processes \citep{proust2019joint}. However, with these focuses, it becomes challenging to infer the underlying disease progression across domains.

These previous works and the updated Jack model motivate us to study the underlying AD progression as a continuous latent variable and jointly model the longitudinally observed AD biomarkers as sigmoid functions of the AD progression, which constitutes a non-linear mixed effect model. Specifically, we assume that a continuous latent disease process affects the multifaceted and longitudinal biomarkers through sigmoid link functions, resulting in the S-shaped biomarker progression trajectories as depicted in the Jack model. We also assume that individual biomarker trajectories (as a function of the underlying AD progression) have covariate-specific intercepts that account for possible heterogeneity among individuals in terms of susceptibility and resistance to pathological changes, such as those due to cognitive reserve and comorbidity. The proposed model is intended to provide data-driven information to help characterize 1) the temporal ordering of biomarker cascades in relation to the AD progression, 2) the multifaceted profile of biomarkers at various stages of AD progression, and 3) disease progression based on biomarker data without relying on clinical diagnoses. These characterizations have the potential of unveiling new insights by going beyond diagnosis-based analyses that can be biased towards existing practice. 

Further, we leverage the proposed model to analyze two longitudinal observational studies, BIOCARD and ADNI (more details in Section \ref{sec:intro-data}). These studies span over two decades and collected longitudinal biomarker measurements since participants  were cognitively normal,  providing a valuable resource for potentially studying the AD biomarker cascade during the preclinical stage. On the other hand, we acknowledge that the slow progression of AD results can result in the preclinical and clinical AD stages span several decades, which may surpassing the duration of current studies.  In addition, practical constraints such as study budgets and the invasive nature of some data collection procedures limit the frequency of biomarker measurement. These issues are currently common in many AD studies with longitudinal biomarker measurements and create challenges in modeling and estimation. We propose modeling and estimation techniques to address these challenges, which have potential applications beyond the specific studies we considered here.

The remainder of this paper is structured as follows. In Section \ref{sec:methodology}, we present the methods, including discussion on general models and special strategies addressing real-world data challenges. In Section \ref{sec:simulation}, we present simulation studies illustrating the empirical performance of proposed methods under various less-than-ideal settings similar to those in the real data. Analysis results in two real-world AD longitudinal follow-up study datasets are presented in Section \ref{sec:data-analysis}, followed by discussions in Section \ref{sec:discussion}.

\section{The BIOCARD and ADNI study}\label{sec:intro-data}
Our methods are motivated by the need to understand biomarker cascades with particular considerations in two datasets: 1) the BIOCARD: Predictors of Cognitive Decline Among Normal Individuals (BIOCARD) \citep{albert2014cognitive} and 2) the subset of cognitively normal participants at baseline in Alzheimer's Disease Neuroimaging Initiative (ADNI) study data \citep{mueller2005alzheimer}. We consider these two datasets because they contain longitudinal data for participants followed from when they were cognitively normal at baseline, have considerably long follow-up duration, and are similar in protocols for biomarker measurements. 

The BIOCARD study is a longitudinal and observational study initiated at the National Institute of Health in 1995. At baseline, 349 cognitively normal individuals that were primarily middle-aged and had a first-degree relative with dementia were recruited and followed by NIH (1995 to 2005) and Johns Hopkins University (2009 to present). During the active study duration, clinical assessments with cognitive tests were done annually for study participants. Magnetic resonance imaging (MRI), cerebrospinal fluid (CSF), and blood specimens were collected biannually. Study participants also receive consensus diagnoses at annual clinical visits, categorized as either cognitively normal (CN), mild cognitive impairment (MCI), or dementia. For individuals that receive a diagnosis of dementia, the age at clinical symptom onset is determined based on the clinical dementia rating (CDR) interview. Individuals in the BIOCARD study have been followed for an average of 15.5 years.

The ADNI study is a multi-center observational study initiated in 2003. The ADNI study recruited participants across the spectrum of cognitive impairment from the categories of CN, MCI, and dementia, defined based on memory criteria, the Mini-Mental State Examination (MMSE), and CDR score. Participants were evaluated at 6 or 12-month intervals for 2 to 3 years, depending on the clinical diagnosis at baseline, and have ongoing annual follow-ups for clinical, imaging, genetic, and biospecimen data collection for up to 16 years. To make analysis results and interpretations more comparable between the BIOCARD and ADNI data, we focus on participants that were CN at baseline. Our analyses include 2115 participants that were CN at baseline with at least one biomarker measurement. The average follow-up time for this cohort is 3.6 years (range: 0–15 years). 

\section{Methodology}\label{sec:methodology}

\subsection{Non-linear mixed effects model for the biomarker disease model}\label{sec:nlme}

We consider a set of observations for $N$ independent participants and $K$ distinct longitudinal biomarkers. Let $Y_{ijk}$ denote the value of the $k$th biomarker for the $i$th participant at the $j$th visit, where $i=1, \dots, N$, $j=1, \dots, J_{i}$, $k\in \kappa_{ij}\subset \big\{1, \dots, K\big\}$. Denote by $D_{ij}$ a continuous latent variable quantifying the underlying AD disease progression, for the $i$th participant at time point $t_{ij}$. 

Following the description in \cite{jack2013update}, we model biomarker levels as sigmoid-shaped functions of the AD disease progression as follows: 
\begin{equation}\label{eq:model-biomarker}
Y_{ijk} = \bX_{ij}^T \bbeta_{k} + f_{k} (D_{ij}; \bgamma_{k}) + \epsilon_{ijk},
\end{equation}
where $\bX_{ij}$ is a $p$-dimensional vector denoting the $i$th participant's covariate information (including an intercept) at the $j$th visit, and $\bbeta_{k}$ is the $p$-dimensional biomarker-specific regression coefficient. In addition to the correlation through the shared disease progression, we allow biomarkers collected at the same visit to correlate with each other through the error term $\epsilon_{ijk}$. This is particularly relevant when multiple biomarkers are collected from the same domain of interest (e.g., Digit Symbol Substitution Test and Logical Memory delayed recall from the cognitive tests), as correlations likely exist due to the shared measurement construct in addition to the shared underlying disease. Specifically, let $\bepsilon_{ij\cdot} = (\epsilon_{ij1}, \dots, \epsilon_{ijK})^{\T}$. We assume that $\bepsilon_{ij\cdot} \sim N(0, \bSigma_{\epsilon})$, where $\bSigma_{\epsilon} = \bSigma(\sigma_1, \dots, \sigma_K, \bphi^{\T})$ is the variance-covariance matrix parametrized by the marginal variance parameters $\sigma_1, \dots, \sigma_K$ and a vector $\bphi$ depicting the correlation structure.

Without loss of generality, we also assume that the biomarkers have been transformed such that higher biomarker values are associated with greater AD disease progressions, indicating further disease progression. We then use a logistic function to formulate the Jack model's hypothesis on biomarker progression:
\begin{equation}\label{eq:model-Dtransformation}
f_{k} (D_{ij}; \bgamma_{k}) = \frac{\gamma_{k1}}{1+\exp\Big\{-\gamma_{k2}\big(D_{ij} - \gamma_{k3}\big)\Big\}}, \text{ where }\gamma_{k1}, \gamma_{k2}>0 .
\end{equation}
The logistic function is chosen as it covers a range of sigmoid-shapes similar to the curves proposed in the Jack model. The parameters also have intuitive interpretations: $\gamma_{k1}$ is the maximum height of the curve, representing the maximum change towards abnormality for the $k$th biomarker during the AD cascade that we are interested in; the parameter $\gamma_{k2}$ reflects the steepness of the sigmoid curve, representing the speed of biomarker level change during AD progression; and $\gamma_{k3}$ is the reflection point of the sigmoid curve, which can be considered as the location when biomarker changes are halfway towards the highest abnormality values of interest.

The relationship between covariates and AD disease progression $d_{ij}$ is modeled as
\begin{equation}\label{eq:model-underlyingD}
D_{ij} = \bZ_{ij}^{\T} \balpha + \delta_{ij},
\end{equation}
where $\bZ_{ij}\in\mathbb{R}^q$ denotes covariates including demographic variables such as age and risk factors such as ApoE-4 carrier status, and $\balpha\in\mathbb{R}^q$ are coefficients. For model identifiability, we fix the location and scale of the latent variable $D_{ij}$ by removing the intercept term in $\balpha$ and setting the variance of $\delta_{ij}$ to be one. For the remainder of this paper, we will use lowercase notations to denote the data realization that corresponds to random variables.

We estimate the proposed model through likelihood estimation. Let $\Phi_q(\mu, \bSigma)$ denote the distribution function of a $q$-dimensional multivariate normal distribution with mean $\mu$ and variance-covariance matrix $\bSigma$, and let $e_{\kappa}\in \R^K$ denote a vector with ones in elements indexed by index set $\kappa\subset\big\{1,\dots, K \big\}$ and zeros elsewhere. Furthermore, let $\bY_{ij\cdot} = (Y_{ij1}, \dots, Y_{ijK})^{\T}$, $\bbeta = (\bbeta_1^{\T},\dots, \bbeta_K^{\T})^{\T}$, $\btheta = (\bbeta_1^{\T},\dots, \bbeta_K^{\T}, \bgamma_1^{\T}, \dots, \bgamma_K^{\T}, \sigma_1, \dots, \sigma_K, \bphi^{\T})^{\T}$, $f_{\cdot}(D_{ij};\bgamma_1, \dots, \bgamma_K) = \big\{f_1(D_{ij};\bgamma_1), \dots, f_K(D_{ij}; \bgamma_K)\big\}^{\T}$, where $\odot$ denote the element-wise multiplication and $\cdot$ denote the matrix multiplication. 
In contrast to vector $\bY_{ij\cdot}$ containing all biomarker values including those not observed, we denote by $\bY_{ij,\kappa_{ij}}$ the observed biomarker vector containing elements index by $k\in\kappa_{ij}$. 

Longitudinal observations of biomarkers are subject to two layers of missingness: 1) a clinical visit can be entirely missing because the participants did not come in as scheduled; 2) when participants did come in, a subset of biomarkers maybe missing due to failure in performing those assessments or due to study design of less frequent collection for some biomarkers. We assume missingness at random (MAR) for both layers of missingness. MAR assumption is commonly accepted for missingness due to administrative scheduling of biomarker collections, whereas we note that it is a stronger assumption than other types of missingness, but can be reasonable conditional on covariates. The likelihood of the biomarker progression model based on observed biomarker measurements can then be written as 
\begin{align*}
    \begin{split}
        L(\btheta,\balpha) = & \prod_{i = 1}^N \prod_{j = 1}^{J_i} \int_{\R} f_{\bY_{ij,\kappa_{ij}} | D_{ij}} (\by_{ij,\kappa_{ij}} | d, \bx_{ij}, \btheta) \cdot f_{D_{ij} } (d|\bz_{ij}, \balpha)\, \mbox{d} d \\
        =&  \prod_{i = 1}^N \prod_{j = 1}^{J_i}  \int_{\R}\int_{\R^{|\kappa_{ij}|}} d\,\Phi_{|\kappa_{ij}|} \big[\{  y_{ij\cdot} - \bx_{ij}^{\T}\bbeta - f_{\cdot} (d; \bgamma_1, \dots, \bgamma_K)\}\odot e_{\kappa_{ij}},\\
        & e_{\kappa_{ij}}\cdot \bSigma_{\epsilon} \cdot e_{\kappa_{ij}}^{\T} \big]\, d\, \Phi_1(d - \bz_{ij}^{\T}\balpha, 1).
    \end{split}
\end{align*}
where $e_{\kappa_{ij}}$ is a vector indicating which biomarkers are missing from the data for the ith participate at the jth visit ($e_{\kappa_{ijk}}=0$ if the the kth biomarker is missing and $e_{\kappa_{ijk}}=1$ otherwise).
Parameter estimates can be obtained through likelihood maximization. 

\subsection{Modeling considerations with the real data}\label{sec:data-cosideration}
In addition to the proposed estimation based on likelihood, our real data application calls for a few special considerations. These include degenerative cases of the logistic function, determining biomarker abnormality range when data do not cover the full longitudinal span of disease progression, and balancing computational speed and estimation and numerical accuracies, which we discuss in the subsections below.

\subsubsection{Range of biomarker progression of interest given limited longitudinal span}\label{sec:data-consideration-gamma1}
Studies that recruit CN participants at enrollment and follow them over time provide uniquely valuable data for studying biomarker progression during preclinical AD. On the other hand, due to the slow progression of AD, often only a limited portion of the study participants reaches AD during the study period where biomarker measurements are collected. This data structure creates challenges for quantitatively studying the hypothesized biomarker cascades. Specifically, ``late-developing" biomarkers, such as cognitive tests, are unlikely to have been adequately observed during the later plateau portion of the sigmoid curves, causing estimability concerns for the height parameters. These parameters may be weakly estimable, or the estimates may have large bias and variability.

Our simulations showed that, if one can provide the height parameter of abnormality change $\gamma_{k1}$ for the biomarkers that reach abnormality the last in the biomarker cascades, the full biomarker progression curves can be recovered even when observations only cover a portion of the curves towards later stages. This provides a practical option in the real world, as the ``late-developing" markers are those related to cognitive performance, which are also relatively standardized and well studied. On the other hand, even though the height parameter is only required for the ``late-developing" biomarkers for consistent estimates, providing additional height parameters when possible will reduce bias in small sample settings. 

We also note that, with finite samples, using the entire theoretically viable range to declare the height parameter for a biomarker might not be a reasonable modeling strategy, especially in the presence of skewed distributions with extreme outliers. Besides, even for bounded cognitive tests such as the Mini-Mental State Examination (MMSE) taking values between 0 and 30, it is relatively rare to observe scores below 10. Meanwhile, examining biomarker progression beyond this stage may provide little clinical utility. Instead, we propose to focus on the progression range \textit{``of interest''} to increase estimation efficiency. 

With the above considerations, we borrow from empirical knowledge to inform the corresponding $\gamma_{k1}$ values. Specifically, we take the difference between the average biomarker values measured at \textit{cognitively normal} visits and \textit{dementia} visits. In an attempt to estimate the average biomarker values from the ``truly normal'' visits at which the biomarkers are believed to be in the initial slow progressing stage, rather than the ``apparent normal'' visits at which the biomarkers might have started to deteriorate, we use the baseline values of individuals who have at least 5 years follow-up and have stayed cognitively normal throughout the entire follow-up period. Since cognitive tests are relatively standardized, we leverage both the BIOCARD and the ADNI study (including the initially MCI and AD cohorts) and take the maximum of the differences from the two studies as the height parameters. For CSF and MRI markers, we use within-study estimates for the consideration of protocol differences and batch effects. Since these parameters are empirically estimated, we conducted simulations in Section \ref{sec:simulation} to evaluate the impact of noise in these estimates on the performance of the proposed model. 

\subsubsection{Degenerative cases of the biomarker progression curve}\label{sec:data-consideration-penalty}
The logistic function used in equation (\ref{eq:model-Dtransformation}) includes degenerative cases that fall outside the hypothesized sigmoid shape when $\gamma_{k2}$ is close to zero or infinity. Specifically, when $\gamma_{k2}\to 0$, we have $f_k(d_{ij}; \bgamma_k) \to  \gamma_{k1}/2$, indicating that the biomarker level is a horizontal line and unrelated to disease progression. When $\gamma_{k2}\to\infty$, we have $f_{k}(d_{ij};\bgamma_k) \to \ind(d_{ij}< \gamma_{k3})\cdot 0 + \ind(d_{ij} > \gamma_{k3})\cdot\gamma_{k1}$, which is a step-function with singularity at $d_{ij} = \gamma_{k3}$, indicating that the biomarker level remains constant over time except for an acute jump when disease progression reaches a certain point. Scientifically, both degenerative cases are unlikely considering evidence in the medical literature. Theoretically, these degenerative cases do not require special handling in ideal situations when we have large sample sizes and large numbers of longitudinal observations. However, in real data where sample sizes and longitudinal follow-ups are limited, pre-hoc exclusion of these degenerative cases can improve estimation stability and reduce the influence of outliers.

To implement this, we restrict the $\gamma_{k2}$ parameters to be between 0 and 10, and add a likelihood penalty term that penalizes extreme values that are close to the 0 and 10 boundaries. We choose to minimize $-\log L +  \lambda\cdot\sum_{k=1}^K n_k^{1/4}\cdot g(\gamma_{k2})$, where $g(\gamma_{k2}) = \log(\gamma_{k2}/10) + \log(1-\gamma_{k2}/10)$. The penalty function  $g(\gamma_{k2})$ was chosen because of its desirable properties. First, it imposes penalties on values deviating from a sigmoid slope of 5, a reasonable pre-hoc value away from the extremes. Second, it imposes symmetric penalties on values that are close to the boundaries of extreme values, e.g., 0.1 and 9.9 receive the same amount of penalization. Third, it imposes large penalties for values that are close to the boundaries (we have $\partial g(\gamma_{k2})/\partial \gamma_{k2}\to\infty$ when $\gamma_{k2}\to 0^+$ or $\gamma_{k2}\to 10^-$) and moderate penalties for values that are away from the boundaries, avoiding the introduction of large biases due to over-penalization. Additionally, we included the scaling term $n_k^{1/4}$ to account for the different sample sizes of biomarkers, while ensuring that the effect of the penalty terms is not driven by the more frequently measured biomarkers. The rate of $1/4$ also ensures that the asymptotic normality of the estimator is unchanged. We choose the tuning parameter $\lambda$ through 10-fold cross-validation, and details are discussed in Sections \ref{sec:simulation} and \ref{sec:data-analysis}.

\subsubsection{Correlation parameter in variance-covariance matrices}
We include correlations among biomarkers measurements at the same visit, and choose to assume independence across time (visits), that is, we have $Cor(\epsilon_{ijk},\epsilon_{ij'k'})=0$, representing independent biomarker measurement errors conditional on the underlying disease progression and covariates across time, and $Cor(\sigma_{ij}, \sigma_{ij'})=0$ for any $j\neq j'$, representing temporal independence in disease within a participant. The former is a standard independent error assumption. The latter assumption is  made to expedite the optimization iteration for the simplified likelihood, allowing for a greater number of iterations and ultimately improving the overall optimization accuracy. As shown by simulation studies in Section \ref{sec:simulation}, ignoring possible disease temporal correlation within a participant minimally impacts estimation under settings that are similar to the observed data structures in BIOCARD and ADNI studies. Detailed discussions on computational methods are provided in Section \ref{sec:estimation-computation}.

\subsection{Model estimation and computational considerations}\label{sec:estimation}
\subsubsection{Point estimates via penalized maximum likelihood}\label{sec:estimation-computation}

To obtain parameter estimates, we maximize the penalized likelihood (equivalently, minimize the negative penalized likelihood) using the Newton-Raphson (NR) optimization algorithm. We came across the following three challenges in terms of optimization. 

First, numerical approaches are needed to approximate the likelihood, along with its first-order and second-order derivatives, due to the integration over the underlying disease, involved in the definitions of these quantities. When the integration is multidimensional due to time correlations in the latent variable, Monte-Carlo algorithms can be adopted. On the other hand, if one can assume independence of $d_{ij}$'s across visit time $j$'s, the Gauss-Hermite quadrature (GQ) approach can drastically improve computational speed, which enables a greater number of iterations for higher numerical accuracy in the overall optimization approach \citep{wang2007algorithms}. In our work, we use the GQNR algorithm \citep{pan2003gauss}, a Newton-Raphson algorithm with Gauss-Hermite quadrature approximations, with details presented below. 

Second, considering the Newton-Raphson (NR) optimization algorithm does not guarantee global convergence, we sample starting values for the algorithm from the uniform distribution with possible ranges specified for each parameter. In addition,  we also include as a starting value, the estimates from a simplified model without biomarker correlations. 

Third, at some steps, the Hessian matrix of the negative penalized likelihood may be negative definite, which makes an ascent update direction for the minimization problem. To ensure that the algorithm searches toward the local minimum, we modify the Hessian matrix at each step by adding an identity matrix multiplied by $max(1-\lambda_{min},0)$, where $\lambda_{min}$ is the smallest eigenvalue of the Hessian matrix. This ensures that the Hessian matrices are always positive definite and that the update directions are descending. In addition, since we are dealing with a relatively high-dimensional parameter space compared to the scale of the datasets, numerical stability is a major concern. It is worth noting that modifying the search directions also significantly enhances the numerical stability. 

Next, we present the details of the algorithm for obtaining estimates of $\psi=(\theta^T, \alpha^T)^T$. Equivalent to maximizing the penalized likelihood, we minimize the negative of it:
\begin{align*}
    \min_{\psi}  l^{\prime}(\psi) := -\log L(\psi) +  \lambda\cdot\sum_{k=1}^K n_k^{1/4}\cdot g(\gamma_{k2})
\end{align*}

Updates of the parameter estimates are made using the Newton-Raphson algorithm. Denote by $\psi^{(m)}$ the parameter estimate at the $m$th step of Newton-Raphson iterative algorithm ($m = 1, 2,\dots$). 
 
\begin{equation}\label{eq:NR-algorithm}
\begin{aligned}
\psi^{(m+1)} &= \psi^{(m)} - \Big( H(\bx, \by, \bz, \psi^{(m)}) + max \{1-\lambda_{min}(H(\bx, \by, \bz, \psi^{(m)})),0 \} I \Big) ^{-1} g(\bx, \by, \bz, \psi^{(m)}),\\
\end{aligned} 
  \end{equation}

where I is the identity matrix, and
\begin{align*}
      g(\bx, \by, \bz, \psi)&=\frac{ \partial l^{\prime}(\psi)}{\partial \psi} = -\frac{ \partial \log L(\psi)}{\partial \psi} + \lambda\sum_{k=1}^K n_k^{1/4}\cdot \frac{\partial g(\gamma_{k2})}{\partial \psi} \\
      H(\bx, \by, \bz, \psi) &= \frac{ \partial^2 l^{\prime}(\psi)}{\partial \psi^T \partial \psi} = -\frac{ \partial^2 \log L(\psi)}{\partial \psi^T \partial \psi} + \lambda\sum_{k=1}^K n_k^{1/4}\cdot \frac{\partial^2  g(\gamma_{k2})}{\partial \psi^T \partial \psi} \\
\frac{ \partial \log L(\psi)}{\partial \psi} &=\sum_{i = 1}^N \sum_{j = 1}^{J_i} \E_{D_{ij} | \bY_{ij,\kappa_{ij}},\psi}  \Big[U_{ij}(\psi) \big| \by_{ij}, \psi \Big]
\end{align*}
\begin{align}\label{e:hessian}
\frac{ \partial^2 \log L(\psi)}{\partial \psi^T \partial \psi}  &= \sum_{i = 1}^N \sum_{j = 1}^{J_i} \E_{D_{ij} | \bY_{ij,\kappa_{ij}},\psi}  \Big[I_{ij}(\psi) \big| \by_{ij}, \psi \Big]+ \sum_{i = 1}^N \sum_{j = 1}^{J_i} \E_{D_{ij} | \bY_{ij,\kappa_{ij}},\psi} \big[U_{ij}(\psi)U_{ij}^T(\psi)\big| \by_{ij}, \psi \big] 
\end{align}
\begin{align*}
&- \sum_{i = 1}^N \sum_{j = 1}^{J_i} \E_{D_{ij} | \bY_{ij,\kappa_{ij}},\psi} \big[U_{ij}(\psi)\big| \by_{ij}, \psi \big]\E_{D_{ij} | \bY_{ij,\kappa_{ij}},\psi} \big[U_{ij}^T(\psi)\big| \by_{ij}, \psi \big] \\
  U_{ij}(\psi) & = \Big(\frac{\partial \log f_{\bY_{ij,\kappa_{ij}} | D_{ij}} (\by_{ij,\kappa_{ij}}|D_{ij}, \bx_{ij}, \btheta) }{\partial \btheta} , \frac{\partial \log f_{D_{ij}} (D_{ij}|\bz_{ij},\balpha)}{\partial \balpha}\Big)^T \\
  I_{ij}(\psi) &= \frac{\partial U_{ij}(\psi)}{\partial \psi}
\end{align*}

 The term $\partial^2 L(\psi)/\partial \psi^T \partial \psi$ is computed using the Louis' formula \citep{louis1982finding}. An interesting observation is that implementing the Newton-Raphson update is marginally equivalent to the EM algorithm update with a one-step Newton-Raphson in the M-step \citep{wang2007algorithms}. %{\red cite Jing Wang paper.}

During the iterative updating of the parameter estimates, closed-form expressions 
\begin{align}\label{eq:NR-component}
    \sum_{i=1}^N\sum_{j = 1}^{J_i}\E_{D_{ij}|\bY_{ij,\kappa_{ij}}, (\btheta, \balpha) = (\btheta^{(m)}, \balpha^{(m)})}\big[ h(D_{ij}) \big| \by_{ij,\kappa_{ij}} \big],
\end{align}
in (\ref{eq:NR-algorithm}) are unavailable, where $h(D_{ij})$ is some function of $D_{ij}$. In our case, specific forms of $h(D_{ij})$ are $\big\{U_{ij}(\psi)\big\}\big|_{\psi = \psi^{(m)}}$,$\big\{I_{ij}(\psi)\big\}\big|_{\psi = \psi^{(m)}}$,,and $\big\{U_{ij}(\psi)U_{ij}^T(\psi)\big\}\big|_{\psi = \psi^{(m)}}$ at the $m$th step, respectively. To compute (\ref{eq:NR-component}), we use the Gauss-Hermite quadrature approximation—an approximation to $\E_{D_{ij}| \bY_{ij,\kappa_{ij}}}[h(D_{ij})|\by_{ij,\kappa_{ij}}]$ via Q-point Gauss-Hermite quadrature 
\begin{align}\label{eq:Q-point-GH}
    \begin{split}
        \E_{D_{ij}| \bY_{ij,\kappa_{ij}}}[h(D_{ij})|\by_{ij,\kappa_{ij}}] &= \frac{\int_{-\infty}^{\infty} h(d) f_{\bY_{ij,\kappa_{ij}}|D_{ij}} (\by_{ij,\kappa_{ij}} | d, \bx_{ij},\btheta) f_{D_{ij}}(d|\bz_{ij},\balpha) \,\mbox{d} d }{\int_{-\infty}^{\infty} f_{\bY_{ij,\kappa_{ij}}|D_{ij}} (\by_{ij,\kappa_{ij}}|d, \bx_{ij}, \btheta) f_{D_{ij}} (d|\bz_{ij},\balpha)\,\mbox{d}d }\\
    &\approx \frac{\sum_{q = 1}^Q w_q\cdot h(2^{1/2}b_q+ \bz_{ij}^{\T} \balpha)f_{\bY_{ij,\kappa_{ij} |D_{ij}}} (\by_{ij,\kappa_{ij}}|d = 2^{1/2} b_q+ \bz_{ij}^{\T} \balpha,\bx_{ij}, \btheta)}{\sum_{q = 1}^Q w_q \cdot f_{\bY_{ij,\kappa_{ij}}|D_{ij}} (\by_{ij,\kappa_{ij}}|d =2^{1/2} b_q+ \bz_{ij}^{\T} \balpha, \bx_{ij},\btheta)},
    \end{split}
\end{align}
where $b_q$ is the $q$-th root of the physicists' version of the Hermite polynomial, and $w_q$ is the associated weight.

\subsubsection{Variance-covariance estimation through Louis' formula}
The variance-covariance matrix of the coefficient estimators can be estimated by the inverse of the observed information matrix and calculated using the Louis' formula \citep{louis1982finding} as $I(\hat{\psi}) = \frac{ \partial^2 \log L(\psi)}{\partial \psi^T \partial \psi}\Big|_{\psi=\hat{\psi}}$, which is the same quantity in equaltion~\ref{e:hessian} evaluated at $\hat{\psi}=(\hat{\btheta}^T,\hat\balpha^T)^T$, the final estimate of the penalized maximum likelihood. 

Louis' formula is applicable to our estimation because the Newton-Raphson algorithm described in Section \ref{sec:estimation-computation} is equivalent to a one-step EM algorithm. Furthermore, all components needed for applying the Louis' formula are calculated in the last step of Newton-Raphson update. In addition, we use the generalized Cholesky decomposition, as implemented in R package \texttt{matrixcalc}, for the observed information matrix to address the potential issue of singular or non-invertible information matrix due to numerical variation. A generalized inversion of matrix can then be computed and used as the estimator for $\mbox{Var}(\widehat{\btheta}, \widehat{\balpha})$. 

\subsection{Prediction and monitoring of disease and biomarker trajectories}\label{sec:method-prediction}
Given model estimation results, we can use participants' observed biomarker information to infer and project their underlying disease $D_i$ at the current visit and a future time. We can also use biomarker distribution among participants with similar underlying diseases to detect potential deviation from ``normal" biomarker trajectories and alert potentially unusual deterioration. 

We first introduce some notations for this section. Let $\widehat\btheta = (\widehat\bbeta_1^{\T}, \dots, \widehat\bbeta_K^{\T}, \widehat\bgamma_1^{\T}, \dots, \widehat\bgamma_K^{\T}, \widehat\sigma_1, \dots, \widehat\sigma_K, \widehat\bphi^{\T})^{\T}$ and $\widehat\balpha$ denote the parameter estimates. Denote by $\bZ_{ij} = \bz(t_0)$ and $\bX_{ij}=\bx(t_0)$ the covariates collected at $t_0$ for the $i$th participant at the $j$th visit, and by $\bz(t)$ and $\bx(t)$ the corresponding covariates at a general time point $t$. In our  application, most components of $\bX_{ij}$ and $\bZ_{ij}$ are time-invariant variables such as ApoE-4 carrier status, gender, and years of education. The time-varying variable in both $\bX_{ij}$ and $\bZ_{ij}$ is age, which is known for any specified future time. We require all covariates to be observed, but allow for missingness in biomarker measurements. 

\subsubsection{Disease progression estimation and prediction}\label{sec:method-prediction-disease}
We synthesize information on covariates and biomarkers to predict the underlying disease progression, either for observed visits given covariates and observed biomarkers, or for future time points given covariates and extrapolated biomarkers values. The extrapolation is calculated by assuming that the biomarker deteriorates according to the normal aging course described in equation (\ref{eq:model-biomarker}), that is, $\widetilde\by_{\kappa}(t) = \by_{\kappa}(t_0) + \big\{ \bx(t)-\bx(t_0) \big\}^{\T}\widehat\bbeta_{\kappa}$. As a special case, we have $\widetilde\by_{\kappa}(t_0) = \by_{\kappa}(t_0)$.

 Conditional on covariates $\bX_{ij} = \bx(t)$ and $\bZ_{ij} = \bz(t)$, the estimated the joint probability density function for a biomarker vector $\bY_{ij\kappa}$ containing observed measurements (missing measurements are integrated out) and disease status is $\hat f_{\bY_{ij\kappa}, D_{ij}|\bX_{ij},\bZ_{ij}}(\by_{\kappa}, d|\bX_{ij}=\bx(t), \bZ_{ij}=\bz(t))
    = \phi_{|\kappa
|}\big[\big\{ \by_{\cdot}-\bx(t)^{\T}\hat\bbeta - f(d;\hat\bgamma_k)\big\}\odot \be_{\kappa}\cdot
    \{\be_{\kappa}\cdot\hat\bSigma \cdot \be_{\kappa}^{\T}\}^{1/2} \big] \cdot\phi_1\big\{ d-\bz(t)^T\balpha \big\}$, where $\phi_{k}$ is the probability density function for a $k$-dimensional standard normal distribution. Based on this joint distribution, we can derive the conditional distribution of disease given covariates $\bx(t)$, $\bz(t)$ and the observed or extrapolated biomarker values $\widetilde\by_{\kappa}(t)$ as 
$\hat f_{D_{ij}|\bY_{ij,\kappa}, \bX_{ij}, \bZ_{ij}}(d|\widetilde\by_{\kappa}(t),\bx(t), \bz(t)) = \hat f_{\bY_{ij,\kappa}, D_{ij}|\bX_{ij}, \bZ_{ij}}(\widetilde\by_{\kappa}(t), d|\bx(t), \bz(t)) / \hat f_{\bY_{ij,\kappa}|\bX_{ij}, \bZ_{ij}}(\widetilde\by_{\kappa}(t)|\bx(t),\bz(t))$. We use GQ approximation to compute the conditional expectation of $D_{ij}$ given covariates and biomarkers, and draw Monte Carlo sample $\big\{ d^{(\ell)} \big\}_{\ell = 1}^L$ from the conditional distribution to construct prediction intervals. For the Monte Carlo sampling, we adopt the rejection sampling algorithm \citep{neal2003slice}, which samples independently $D$ from some candidate distribution $p(d)$ and $U$ from a uniform distribution and checks whether $U<\hat f_{D_{ij}|\bY_{ij,\kappa}, \bX_{ij}, \bZ_{ij}}(d|\widetilde\by_{\kappa}(t),\bx(t), \bz(t))/M\cdot p(d)$. Specifically, we use normal candidate distributions. To improve the efficiency of rejection sampling, we adjust the candidate distributions for each subject of prediction interest to be close to the target distribution—we set the mean of the normal distribution to be the expectation of the target distribution and use a standard deviation of 10 to bound the tails. Constant $M$ is chosen by finding the maximum of $\hat f_{D_{ij}|\bY_{ij,\kappa}, \bX_{ij}, \bZ_{ij}}(d|\widetilde\by_{\kappa}(t),\bx(t), \bz(t))/p(d)$ through grid search. The asymptotic variance and confidence interval of conditional expectation of $D_{ij}$ can be computed by simulating from the asymptotic distribution of $(\widehat\btheta,\widehat\balpha)$ and utilizing formula (\ref{eq:Q-point-GH}).

Since the biomarker values are extrapolated with only the deterioration due to the course of normal aging described by the age regression coefficient, this prediction reflects a progression under ``stable disease'' situation. It can be used as a benchmark for detecting potentially more aggressive disease progression, or, as we discuss in more detail below, for detecting biomarker values that are worse than expected under this disease stage, which may also signal a potentially more aggressive disease progression.

\subsubsection{Monitoring and predicting biomarker deterioration}\label{sec:method-prediction-biomarker}
For a time point $t_0$ where biomarkers are measured, we may be interested in the distribution of biomarker values among individuals with the same underlying disease status and covariate values. This can serve as a monitoring tool to detect potential biomarkers that appear to have deteriorated beyond the ``regular'' range in the population with the same disease and covariate profiles. For example, a particular area shown on MRI or in a particular cognitive function may appear to be worse than participants with similar covariates and underlying disease. Since AD progression is a multi-faceted process as discussed in \cite{sperling2011toward}, this biomarker monitoring tool complements the one-dimensional information given in the predicted AD progressions, by potentially revealing a domain worth special attention or signaling the start of accelerated worsening of the underlying disease. 

For a future time point $t_1$ where no data is yet observed, we may be interested in predicting biomarker values under extrapolated AD disease progression described in (\ref{eq:model-underlyingD}), calculated as $\widetilde d(t_1) = d_0 + \big\{ \bz(t_1)-\bz(t_0) \big\}^{\T}\widehat\balpha$, or under other disease progression status of interest. For example, we may be interested in evaluating the plausible future levels of participant's cognitive function, to facilitate appropriate treatment and care plans. 

For either a time point $t_0$ at which biomarker measurements are collected or a future time point $t_1$ at which no data is yet observed, we predict the biomarker distributions given covariates and the extrapolated or specified AD disease risk score $\widetilde d(t)$ ($t = t_0, t_1$), following approaches similar to those discussed in Section \ref{sec:method-prediction-disease}. Specifically, we predict $y_{ijk}$ for given $D_{ij} = \widetilde d(t)$ by simulating, for $\ell = 1,\dots, L$, the error term $\bepsilon_{ij}^{(\ell)} = (\epsilon_{ij1}^{(\ell)}, \dots, \epsilon_{ijK}^{(\ell)}) $ from $K$-dimensional multivariate normal distribution with mean zero and variance-covariance matrix $\hat\bSigma$, and then compute $y_{ijk}^{(\ell)}\{\widetilde d(t)\}  = \bX_{ij}^{\T}\hat\bbeta_k + f_k\{\widetilde d(t);\hat\bgamma_k\} + \epsilon_{ijk}^{(\ell)}$ as a function of $\widetilde d(t)$. The mean of the computed $\big\{ y_{ijk}^{(\ell)}\{\widetilde d(t)\} \big\}_{\ell = 1}^L$ and the $a$\% and $(100-a)$\% quantiles then serve to establish monitoring and prediction tools.

\section{Analyses on biomarker cascading model using the BIOCARD and ADNI data}\label{sec:data-analysis}

\begin{figure}
    \centering
    \includegraphics[width=0.92\linewidth]{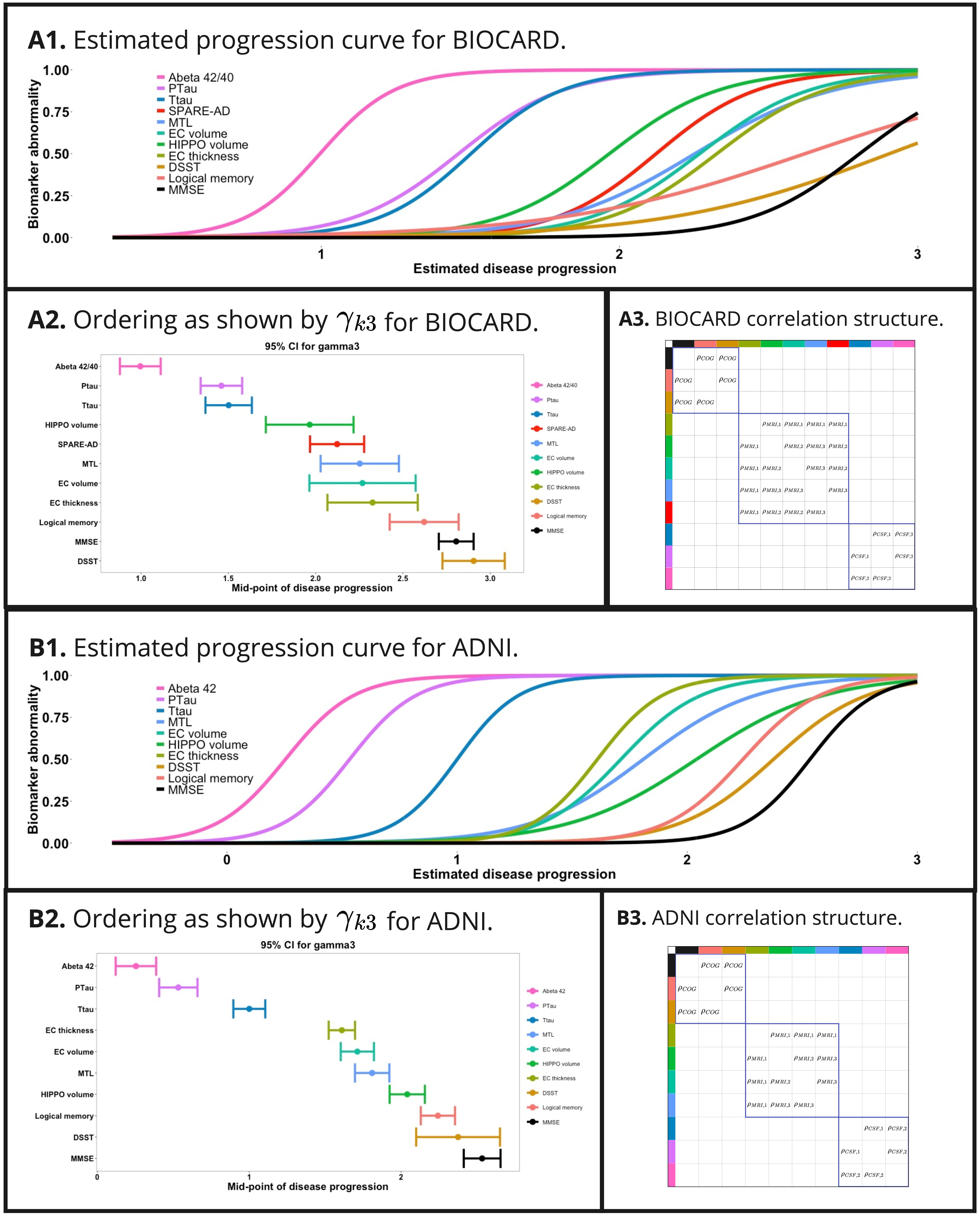}
    \caption{Analysis results for BIOCARD and ADNI data. Panels A1-A3 show results for BIOCARD; panels B1-B3 show results for ADNI. Panels A1 and B1 illustrate the estimated biomarker progression curves $f_k(d)$; panels A2 and B2 illustrate the ordering as shown by the estimated location parameter $\gamma_{k3}$; panels A3 and B3 illustrate the correlation structures for biomarker error terms.}
    \label{fig:data-analysis-result1}
\end{figure}

\begin{figure}
    \centering
    \includegraphics[width=0.82\linewidth]{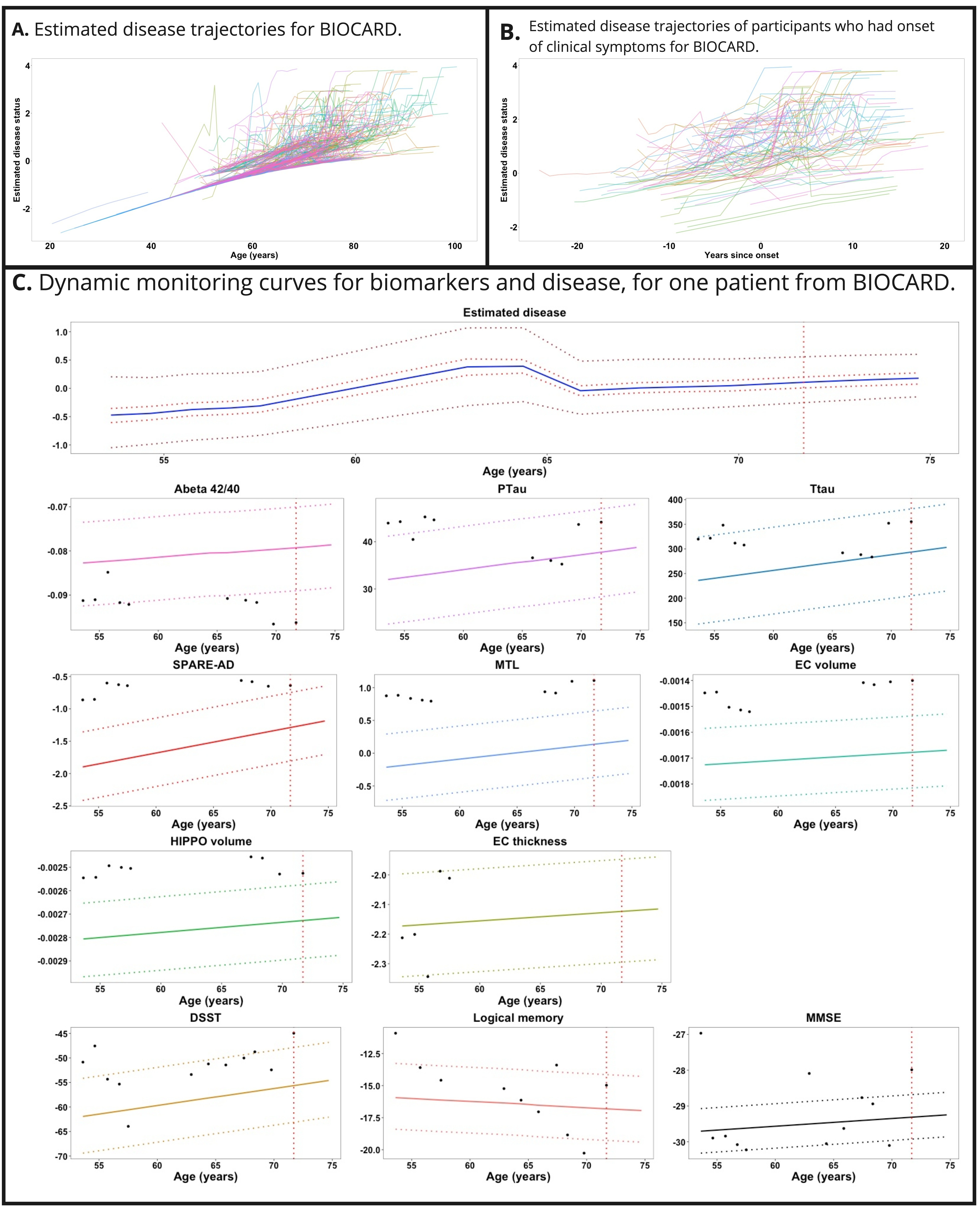}
    \caption{Illustration of predicted disease and biomarker trajectories for the BIOCARD data. Panel A shows the estimated disease trajectories for all participants on the age scale; panel B shows the estimated disease trajectories for participants who had onset of clinical symptoms, with trajectories aligned at the onset age (time zero); panel C illustrates, for one participant from BIOCARD, the disease trajectory with 95\% confidence interval and 50\% prediction interval, and the estimated and predicted biomarker trajectories with 50\% prediction intervals.}
    \label{fig:data-analysis-result2}
\end{figure}

In this section, we present details of the real data analysis and results. We first discuss the considerations and process regarding unbalanced visits and biomarker measurements. We then present the model specifications including covariates and biomarker correlation structures. After detailing the analytic approach, we present estimation results in Tables \ref{table:analysis-BIOCARD}, \ref{table:analysis-ADNI} and Figure \ref{fig:data-analysis-result1}, and showcase prediction and monitoring methods with Figure \ref{fig:data-analysis-result2}.

To create datasets suitable for analyses, we merge biomarker measurements to harmonize the irregular visit times. Out of considerations for model misspecification as discussed in Section \ref{sec:data-cosideration}, we use ``anchoring visits" to which we merge other biomarker measurements. Specifically, for the BIOCARD data, we use cognitive visit dates as the anchoring visit date, whereas for the ADNI data, we generate annually spaced dates starting from baseline as cognitive visit dates, to merge various cognitive test scores collected over a few days for the same visit. Next, for both the BIOCARD and ADNI data, we merge MRI and CSF biomarker measurements taken within $\pm$ two-year windows of an anchoring visit onto the anchoring visit date. When multiple biomarker measurements exist within the measurement window, we merge the temporally closest measurement. We allow a single biomarker measurement to be used multiple times in merging if it is the closest in multiple windows. We use a $\pm$ two-year window for merging the MRI and CSF biomarker measurements because these biomarkers are less frequently collected but are considered to be reasonably stable within a two-year window. For merging cognitive visits in ADNI to the annually spaced anchoring dates, we use a $\pm$ one-year window.

We consider the modeling of the following biomarkers in data analyses: A$\beta$ (A$\beta$ 42/40 ratio for the BIOCARD, A$\beta$ 42 for ADNI), phosphorylated tau (PTau) and total tau (Ttau) from the CSF domain; entorhinal cortex (EC) thickness, EC volume, hippocampal (HIPPO) volume, medial temporal lobe composite score (MTL), which is constructed based on volumes of HIPPO, EC and amygdala \citep{soldan2015relationship}. Further, we include the Spatial Pattern of Abnormality for Recognition of Early Alzheimer’s Disease score (SPARE-AD; BIOCARD only) \citep{davatzikos2009longitudinal} from the MRI domain; and the Digit Symbol Substitution Test (DSST) score, the Mini-Mental State Exam (MMSE) score, and the Logical Memory delayed recall (LM) score from the cognitive test domain. We include in the biomarker model (\ref{eq:model-biomarker}) age, gender, and years of education as covariates, and we include in the AD disease progression model (\ref{eq:model-underlyingD}) age and ApoE-4 carrier status as covariates. We assume temporal independence across visits, and assume biomarkers model errors to be correlated within the same domain and independent between domains. Specifically, we specify the error correlation structure according to preliminary data analyses on the marginal correlations between markers. Detailed correlation structures are shown in Figures \ref{fig:data-analysis-result1}.A3 and \ref{fig:data-analysis-result1}.B3.

As discussed in Section \ref{sec:data-cosideration}, we have a few additional data considerations during estimation. First, we predetermine the values of $\gamma_{k1}$ from observed biomarker values as discussed in \ref{sec:data-consideration-gamma1}. Second, we incorporate the penalty term as described in Section \ref{sec:data-consideration-penalty}, choosing the tuning parameter $\lambda$ from 16 candidate values between $10^{-2}$ and $10^2$ according to 10-fold cross validations. The tuning parameter is chosen to be $10^{0.3}\approx 2$ for both data. Coefficient estimates with 95\% confidence intervals are presented in Table \ref{table:analysis-BIOCARD} for BIOCARD and in Table \ref{table:analysis-ADNI} for ADNI. We further illustrate in Figure \ref{fig:data-analysis-result1} the estimated biomarker progression curves and biomarker cascade ordering as shown by the $\gamma_{k3}$ parameters. After the model estimation, we carry out prediction of the disease and biomarker trajectories as described in Section \ref{sec:method-prediction}, and we show results in Figure \ref{fig:data-analysis-result2} as an illustration.

We observe from analysis results, especially the biomarker progression curves as shown in Figure \ref{fig:data-analysis-result1}, that the BIOCARD and ADNI data exhibit similar patterns: biomarkers from the CSF domain are the first to deteriorate, followed by those from the MRI domain and then the cognitive domain. In particular, within the CSF domain, A$\beta$ is consistently shown to progress first. The ordering of biomarkers within the MRI and cognitive domains are less conclusive due to wide confidence intervals. Meanwhile, we observe that the biomarker progression curves cross at some levels, which implies that the biomarker cascade ordering could be different depending on the level of biomarker abnormality (y-axis). Of note, the disease scales in BIOCARD and ADNI data, as reflected by the x-axes in Figures \ref{fig:data-analysis-result1}.A1 and \ref{fig:data-analysis-result1}.B1, are different because both biomarkers and covariates (standardized to have mean zero and standard deviation one if continuous) are on different scales.

To illustrate disease prediction and biomarker monitoring procedures discussed in Section \ref{sec:method-prediction}, we show in Figure \ref{fig:data-analysis-result2}.A the estimated disease trajectories (given covariates and available biomarkers) over age, and in Figure \ref{fig:data-analysis-result2}.B the estimated disease trajectories along a timescale relative to the time of onset, for participants who had onset of clinical symptoms. In Figure \ref{fig:data-analysis-result2}.C, we show an example of a biomarker and disease monitoring panel using longitudinal data from one participant, with the red dotted vertical line separating observed and future visits. For each biomarker, we present the predicted biomarker values given covariates and the estimated or projected disease for the observed or future visits, accompanied by dotted lines showing the upper and lower boundaries of the 50\% prediction intervals. We add observed biomarker values (with noise added to protect participant privacy) as discrete points in the plot to see the relative positions of the observed values in reference to the predicted biomarker profile, among a population similar in terms of age, gender, education, ApoE-4 carrier status, and underlying disease status. In addition to the biomarkers, we also show the estimated and projected disease trajectory of that participant, accompanied by a 95\% CI (the narrower interval) and a 50\% prediction interval (the wider interval).

\begin{minipage}{\linewidth}
\centering
\captionof{table}{Parameter estimates based on BIOCARD data.}
\label{table:analysis-BIOCARD}
\resizebox{\linewidth}{!}{
\begin{tabular}[h]{cccccccccc}
\hline\\[-0.8em]
                               & \multicolumn{9}{c}{Progression Curve Parameters}                                                                                                                                                      \\ \hline\\[-0.8em]
                               & \multicolumn{9}{c}{}                                                                                                                                                                                  \\[-1.3em]
                               & Intercept            & Age                  & ApoE-4 carrier      & Gender female        & Education            & $\gamma_{k1}^*$ (fixed) & $\gamma_{k2}$     & $\gamma_{k3}$     & $\sigma_k$        \\ \cline{2-10} \\[-0.8em]
                               & \multicolumn{9}{c}{}                                                                                                                                                                                  \\[-1em]
A$\beta$ 42/40 ratio                       & -0.48 (-0.54, -0.42) & 0.09 (0.06, 0.12)    & 0.60 (0.52, 0.67)    & -0.02 (-0.09, 0.05)  & 0.00 (-0.04, 0.03)      & 1.32                    & 7.99 (6.65, 8.88) & 1.00 (0.88, 1.11)    & 0.64 (0.61, 0.67) \\
pTau                           & -0.35 (-0.41, -0.29) & 0.15 (0.12, 0.19)    & 0.08 (0.01, 0.16)   & 0.07 (0.01, 0.14)    & 0.01 (-0.02, 0.04)   & 2.32                    & 5.67 (4.61, 6.67) & 1.46 (1.34, 1.58) & 0.63 (0.60, 0.65)  \\
tTau                           & -0.24 (-0.30, -0.17)  & 0.20 (0.16, 0.24)     & 0.02 (-0.06, 0.10)   & 0.06 (-0.01, 0.14)   & -0.04 (-0.07, 0.00)     & 1.81                    & 6.49 (4.51, 8.06)  & 1.50 (1.37, 1.63)  & 0.76 (0.73, 0.79) \\
SPARE-AD composite score       & -0.09 (-0.17, -0.02) & 0.40 (0.36, 0.45)     & -0.10 (-0.19, -0.01) & 0.16 (0.07, 0.25)    & 0.09 (0.05, 0.13)    & 1.60                     & 5.89 (3.61, 7.85) & 2.12 (1.97, 2.28) & 0.87 (0.84, 0.90)    \\
MTL composite score            & -0.21 (-0.29, -0.13) & 0.26 (0.21, 0.31)    & -0.10 (-0.20, -0.01)  & 0.38 (0.29, 0.47)    & 0.07 (0.03, 0.12)    & 1.08                    & 4.27 (1.64, 7.38)  & 2.25 (2.03, 2.48) & 0.92 (0.89, 0.95) \\
EC volume$^{\dagger}$          & -0.05 (-0.12, 0.03)  & 0.13 (0.08, 0.18)    & -0.10 (-0.20, -0.01)  & 0.11 (0.02, 0.21)    & -0.01 (-0.06, 0.03)  & 0.70                     & 5.39 (0.79, 9.41)    & 2.27 (1.96, 2.57) & 0.95 (0.92, 0.98) \\
HIPPO volume$^{\dagger}$ & 0.24 (0.17, 0.32)    & 0.18 (0.14, 0.23)    & -0.03 (-0.12, 0.06) & -0.42 (-0.51, -0.33) & 0.10 (0.05, 0.14)     & 0.74                    & 5.42 (1.29, 9.04)  & 1.97 (1.71, 2.22) & 0.89 (0.86, 0.92) \\
EC thickness                   & -0.15 (-0.26, -0.04) & 0.11 (0.04, 0.18)    & -0.04 (-0.16, 0.08) & 0.34 (0.22, 0.46)    & 0.11 (0.05, 0.17)    & 1.58                    & 5.44 (2.08, 8.44) & 2.33 (2.07, 2.58) & 0.96 (0.92, 1.00)  \\
DSST                          & 0.20 (0.14, 0.26)     & 0.29 (0.25, 0.33)    & 0.02 (-0.05, 0.09)  & -0.44 (-0.50, -0.37)  & -0.17 (-0.21, -0.14) & 2.43                    & 2.69 (1.47, 4.40) & 2.90 (2.73, 3.08)  & 0.89 (0.86, 0.91) \\
Logical Memory                           & 0.10 (0.04, 0.17)     & -0.13 (-0.17, -0.08) & -0.01 (-0.09, 0.06) & -0.38 (-0.45, -0.31) & -0.10 (-0.14, -0.07)  & 2.83                    & 2.41 (1.82, 3.12) & 2.62 (2.42, 2.82) & 0.92 (0.90, 0.95)  \\
MMSE                           & -0.04 (-0.08, 0.01)  & 0.16 (0.14, 0.19)    & -0.06 (-0.11, 0.00)    & -0.11 (-0.16, -0.06) & -0.09 (-0.12, -0.07) & 5.87                    & 5.45 (4.56, 6.31)  & 2.80 (2.71, 2.90)   & 0.67 (0.65, 0.69) \\ \hline\\[-0.8em]
                               & \multicolumn{9}{c}{}                                                                                                                                                                                  \\[-1.3em]
                               & \multicolumn{9}{c}{Other Parameters}                                                                                                                                                                  \\ \hline
                               & \multicolumn{9}{c}{}                                                                                                                                                                                  \\[-1.2em]
                               & Age                  & ApoE-4 carrier       &                     & $\rho_{COG}$         & $\rho_{MRI,1}$       & $\rho_{MRI,2}$          & $\rho_{MRI,3}$    & $\rho_{CSF,1}$    & $\rho_{CSF,2}$    \\ \cline{2-3} \cline{5-10} 
                               &                      &                      &                     &                      &                      &                         &                   &                   &                   \\[-0.8em]
Disease model                  & 0.72 (0.70, 0.75)     & 0.59 (0.54, 0.64)    & Correlation         & 0.15 (0.12, 0.18)    & 0.07 (0.05, 0.08)    & 0.45 (0.42, 0.48)       & 0.72 (0.67, 0.77) & 0.70 (0.63, 0.78)  & 0.10 (-0.04, 0.24) \\ \hline
\end{tabular}}
\begin{flushleft}
    \tiny{*: $\gamma_{k1}$ values are predetermined.\\
$\dagger$: MRI volumes are adjusted for intracranial volume by division. Continuous variables were standardized to have mean zero and standard deviation one before model fitting.
}
\end{flushleft}
\end{minipage}

\begin{minipage}{\linewidth}
\centering
\captionof{table}{Parameter estimates based on ADNI data.}
\label{table:analysis-ADNI}
\resizebox{\linewidth}{!}{
\begin{tabular}{cccccccccc}
\hline\\[-0.8em]
                               & \multicolumn{9}{c}{Progression Curve Parameters}                                                                                                                                                        \\ \hline
                               & \multicolumn{9}{c}{}                                                                                                                                                                                    \\[-0.8em]
                               & Intercept            & Age                 & ApoE-4 carrier       & Gender female        & Education            & $\gamma_{k1}^*$ (fixed) & $\gamma_{k2}$     & $\gamma_{k3}$     & $\sigma_k$          \\ \cline{2-10} 
                               & \multicolumn{9}{c}{}                                                                                                                                                                                    \\[-0.8em]
A$\beta$ 42            & -0.69 (-0.77, -0.61) & 0.02 (-0.02, 0.07)  & 0.67 (0.56, 0.77)    & 0.03 (-0.06, 0.12)   & -0.09 (-0.13, -0.05) & 1.24                    & 6.74 (4.47, 8.40)  & 0.25 (0.12, 0.39) & 0.72 (0.68, 0.76)    \\
pTau                           & -0.46 (-0.54, -0.38) & 0.04 (-0.01, 0.10)   & 0.24 (0.13, 0.35)    & 0.13 (0.03, 0.22)    & 0.01 (-0.03, 0.06)   & 1.08                    & 6.93 (4.64, 8.54) & 0.53 (0.41, 0.66) & 0.86 (0.83, 0.90)    \\
tTau                           & -0.40 (-0.47, -0.32)  & 0.09 (0.03, 0.14)   & 0.14 (0.03, 0.25)    & 0.07 (-0.02, 0.16)   & -0.05 (-0.09, -0.01) & 1.79                    & 7.50 (5.71, 8.71)  & 1.00 (0.90, 1.11)     & 0.79 (0.75, 0.83)   \\
MTL composite score            & -0.39 (-0.44, -0.34) & 0.38 (0.35, 0.40)    & 0.05 (-0.01, 0.10)    & 0.37 (0.33, 0.42)    & 0 (-0.03, 0.02)      & 2.35                    & 4.11 (3.02, 5.30) & 1.81 (1.70, 1.92)  & 0.73 (0.71, 0.74)    \\
EC volume$^{\dagger}$          & -0.14 (-0.19, -0.09) & 0.17 (0.14, 0.20)    & 0.00 (-0.06, 0.07)      & -0.03 (-0.08, 0.03)  & 0.01 (-0.01, 0.04)   & 1.85                    & 5.74 (4.10, 7.24) & 1.71 (1.60, 1.82)  & 0.88 (0.86, 0.89)   \\
HIPPO volume$^{\dagger}$ & 0.07 (0.01, 0.12)    & 0.39 (0.36, 0.42)   & 0.03 (-0.02, 0.09)   & -0.35 (-0.4, -0.31)  & 0.08 (0.06, 0.11)    & 2.10                     & 3.63 (1.89, 5.82) & 2.04 (1.92, 2.16) & 0.77 (0.75, 0.78)   \\
EC thickness                   & -0.22 (-0.26, -0.18) & 0.21 (0.19, 0.24)   & -0.09 (-0.15, -0.03) & 0.03 (-0.02, 0.08)   & -0.02 (-0.04, 0.01)  & 2.36                    & 7.09 (5.75, 8.14)  & 1.61 (1.52, 1.70)  & 0.72 (0.70, 0.74)   \\
DSST                        & 0.11 (0.00, 0.22)       & 0.35 (0.25, 0.45)   & 0.09 (-0.07, 0.26)   & -0.44 (-0.58, -0.29) & -0.20 (-0.27, -0.13)  & 2.80                     & 4.93 (0.86, 9.1) & 2.38 (2.10, 2.65)  & 0.93 (0.88, 0.98)   \\
Logical Memory                            & 0.04 (-0.01, 0.09)   & -0.01 (-0.04, 0.02) & 0.02 (-0.05, 0.09)   & -0.27 (-0.34, -0.21) & -0.26 (-0.29, -0.23) & 2.65                    & 6.01 (3.82, 7.86)  & 2.24 (2.13, 2.35) & 0.88 (0.85, 0.90)     \\
MMSE                           & -0.03 (-0.07, 0.01)  & 0.14 (0.11, 0.16)   & 0.07 (0.01, 0.13)    & -0.15 (-0.20, -0.10)   & -0.11 (-0.14, -0.09) & 4.45                    & 7.01 (5.55, 8.14) & 2.53 (2.41, 2.65) & 0.73 (0.71, 0.75)   \\ \hline
                               & \multicolumn{9}{c}{}                                                                                                                                                                                    \\[-0.8em]
                               & \multicolumn{9}{c}{Other Parameters}                                                                                                                                                                    \\ \hline
                               & \multicolumn{9}{c}{}                                                                                                                                                                                    \\[-1em]
                               & Age                  & ApoE-4 carrier      &                      & $\rho_{COG}$         & $\rho_{MRI,1}$       & $\rho_{MRI,2}$          & $\rho_{MRI,3}$    & $\rho_{CSF,1}$    & $\rho_{CSF,2}$      \\ \cline{2-3} \cline{5-10} 
                               &                      &                     &                      &                      &                      &                         &                   &                   &                     \\[-0.8em]
Disease model                  & 0.52 (0.48, 0.55)    & 0.39 (0.36, 0.42)   & Correlation          & 0.19 (0.16, 0.22)    & 0.33 (0.32, 0.35)    & 0.41 (0.37, 0.46)       & 0.67 (0.61, 0.72) & 0.53 (0.47, 0.59) & -0.01 (-0.13, 0.11) \\ \hline
\end{tabular}}
\begin{flushleft}
    \tiny{*: $\gamma_{k1}$ values are predetermined.\\
$\dagger$: MRI volumes are adjusted for intracranial volume by division. Continuous variables were standardized to have mean zero and standard deviation one before model fitting.
}
\end{flushleft}
\end{minipage}

\section{Real-data based simulations}\label{sec:simulation}
In this section, we present a simulation study designed to investigate the finite-sample performance of the proposed methods under scenarios that closely resemble real data, while also examining the robustness of our proposed approach to potential model misspecification.

We emulate real-data structures from the BIOCARD and ADNI datasets in our simulation study by: 1) using observed individual covariates; 2) sampling missingness indicators from the observed missingness structure—specifically, for each individual, we sample the number of visits from the observed number of visits, and then for each visit, we sample the observed status of each biomarker from the real-data observed status of the specific biomarker; 3) generating biomarker outcomes for parameter values that are close to the estimated coefficients from real data analyses. Due to the complexity of our real data analyses, which involve eleven biomarkers, and the accompanying difficulties in presenting simulation results, we simplified the data-generating scenarios to consist of four biomarkers, one of which emulates a cognitive biomarker, two MRI biomarkers, and one CSF biomarker. This design ensures that biomarkers deteriorating at different stages of AD disease progression are represented and investigated in our simulations. Our design of the simulation settings results in sample sizes of 299 for BIOCARD, with an average of 11 visits for each individual, and 2115 for ADNI, with an average of 4 visits for each individual.

To investigate the performance of our proposed approach under correctly specified model assumptions, we consider two scenarios where the data-generating model assumption and the estimation assumption are consistent. In one scenario, model coefficients are estimated with a penalty on \(\gamma_{3k}\)'s; in the other, model coefficients are estimated without any penalty to illuminate the impact of placing penalties on the rate parameters. To understand the impact of potential model misspecification, we consider two additional scenarios where moderate violations of model assumptions are present. We consider a "time correlation" model, where disease errors \(\delta_{ij}\) are assumed to have a correlation of 0.5 across visits within the same individual, aiming to assess how temporal correlation affects the robustness of our method. Additionally, we explore a "perturbed \(\gamma_{1}\)" model, where the sum of a fixed value of \(\gamma_1\) parameters and Gaussian noise (SD is 0.1) is used as the data-generating true \(\gamma_1\)'s, to emulate real-world noise in plugged-in \(\gamma_1\) values and test the sensitivity of our model to such perturbations. For the latter two scenarios, estimation was done with penalties on \(\gamma_{k3}\)'s. These scenarios showcase the robustness of our proposed methods against common and realistic deviations from the assumed model. For each scenario, we simulate 1000 replications and summarize results in Tables \ref{table:simulation-BIOCARD} and \ref{table:simulation-ADNI} in terms of bias, empirical standard error (ESE), mean standard error (MSE), and 95% confidence interval coverage probability (CP).

Results in Tables \ref{table:simulation-BIOCARD} and \ref{table:simulation-ADNI} show that across all simulation scenarios and most parameters, our proposed approach to dealing with data challenges yielded parameter point estimates with small bias, standard error estimates that are on average close to the empirical standard error, and confidence intervals that have proper coverages. We observed some undercoverage in the location parameter \(\gamma_{k2}\) for the second and third biomarkers emulating MRI variables, and some overcoverage in \(\gamma_{k2}\) for the fourth biomarker emulating a CSF variable. However, estimation without penalty terms proved to have much worse stability compared to that with penalty terms. This is not reflected in the standard summary statistics but is apparent in the converged percentage of replications for each simulation scenario—if estimation can be obtained for a replication, that replication is considered converged; if the optimization process for estimation fails to converge for a replication, that replication is considered not converged. For all scenarios where the estimation is done with penalty terms, we observed only occasional failure of convergence, with the convergence percentages all above 98\%. Whereas, for scenarios where estimation is done without penalty terms, only 26.8\% converged for the BIOCARD simulation, and 47.9\% for the ADNI simulation. This result demonstrates the pivotal role of penalty terms in improving the stability and feasibility of estimation given the various data challenges that we faced. Overall, our simulation results show the robust finite-sample performance of the proposed approach in realistic scenarios that allow moderate violations of model assumptions.

% \begin{table}[]
% \centering
% \captionof{table}{Description of simulation scenarios.}
% \label{table:simulation-scenario}
% \begin{tabular}{ll}
% \hline
% Scenario                             & Description \\ \hline
% Correct model, with penalty          &             \\
% Correct model, without penalty       &             \\
% Time correlation model, with penalty &             \\
% Perturbed $\gamma_1$, with penalty   &             \\ \hline
% \end{tabular}
% \end{table}

% \usepackage{multirow}
\begin{minipage}{\linewidth}
\centering
\captionof{table}{Summary statistics for simulations emulating the BIOCARD data.}
\label{table:simulation-BIOCARD}
\resizebox{\linewidth}{!}{
\begin{tabular}{cccccccccccccccccccccc}
\hline\\[-0.8em]
Parameter              & True value           &  & Bias     & ESE   & MSE   & CP &                      & Bias                 & ESE                  & MSE                  & CP                   &  & Bias                 & ESE                  & MSE                  & CP                   &  & Bias                 & ESE                  & MSE                  & CP                   \\ \cline{1-7} \cline{9-12} \cline{14-17} \cline{19-22} 
\multicolumn{1}{l}{}   & \multicolumn{1}{l}{} &  & \multicolumn{4}{c}{Correct model, with penalty}          & \multicolumn{1}{l}{} &  \multicolumn{4}{c}{Correct model, without penalty} &   & \multicolumn{4}{c}{Time correlation model, with penalty} &  & \multicolumn{4}{c}{Perturbed $\gamma_1$, with penalty} \\ \cline{1-2} \cline{4-7} \cline{9-12} \cline{14-17} \cline{19-22} \\[-0.8em]
\multicolumn{1}{c}{Converged percentage}   & \multicolumn{1}{l}{} &  & \multicolumn{4}{c}{99.7\%}          & \multicolumn{1}{l}{} &  \multicolumn{4}{c}{26.8\%} &   & \multicolumn{4}{c}{99.4\%} &  & \multicolumn{4}{c}{99.7\%}\\
$\beta_{Intercept, 1}$ & 0.1                  &  & 0.019    & 0.034 & 0.035 & 93 &                      & -0.0033              & 0.041                & 0.038                & 92                   &  & 0.017                & 0.035                & 0.035                & 92                   &  & 0.019                & 0.034                & 0.035                & 93                   \\
$\beta_{Intercept, 2}$ & -0.15                &  & -0.003   & 0.056 & 0.056 & 95 &                      & -0.027               & 0.087                & 0.077                & 96                   &  & -0.00086             & 0.052                & 0.056                & 97                   &  & -0.0028              & 0.056                & 0.056                & 94                   \\
$\beta_{Intercept, 3}$ & 0.24                 &  & -0.00093 & 0.042 & 0.041 & 94 &                      & -0.01                & 0.044                & 0.049                & 97                   &  & -0.003               & 0.041                & 0.041                & 95                   &  & -0.00062             & 0.042                & 0.041                & 94                   \\
$\beta_{Intercept, 4}$ & -0.35                &  & -0.0037  & 0.035 & 0.035 & 95 &                      & -0.00093             & 0.037                & 0.036                & 94                   &  & -0.0059              & 0.037                & 0.035                & 94                   &  & -0.0051              & 0.038                & 0.035                & 92                   \\
$\beta_{Age, 1}$       & -0.13                &  & 0.011    & 0.022 & 0.022 & 92 &                      & 0.001                & 0.023                & 0.023                & 95                   &  & 0.0098               & 0.022                & 0.022                & 92                   &  & 0.011                & 0.022                & 0.022                & 92                   \\
$\beta_{Age, 2}$       & 0.11                 &  & 0.0006   & 0.035 & 0.035 & 95 &                      & -0.0044              & 0.037                & 0.038                & 96                   &  & -0.0004              & 0.034                & 0.035                & 96                   &  & 0.00065              & 0.035                & 0.035                & 95                   \\
$\beta_{Age, 3}$       & 0.18                 &  & 0.0011   & 0.024 & 0.025 & 95 &                      & -0.0011              & 0.024                & 0.027                & 96                   &  & -0.00021             & 0.025                & 0.025                & 94                   &  & 0.0015               & 0.024                & 0.025                & 95                   \\
$\beta_{Age, 4}$       & 0.15                 &  & 0.00029  & 0.019 & 0.02  & 95 &                      & 0.00041              & 0.019                & 0.02                 & 95                   &  & -0.0014              & 0.02                 & 0.02                 & 95                   &  & -5.9e-05             & 0.021                & 0.02                 & 93                   \\
$\beta_{ApoE, 1}$      & -0.013               &  & 0.0069   & 0.04  & 0.041 & 95 &                      & -0.0048              & 0.042                & 0.043                & 95                   &  & 0.012                & 0.043                & 0.041                & 93                   &  & 0.0069               & 0.04                 & 0.041                & 95                   \\
$\beta_{ApoE, 2}$      & -0.044               &  & 0.00026  & 0.062 & 0.062 & 95 &                      & -0.0035              & 0.066                & 0.064                & 95                   &  & 0.0019               & 0.06                 & 0.062                & 95                   &  & 0.00044              & 0.063                & 0.063                & 95                   \\
$\beta_{ApoE, 3}$      & -0.033               &  & -0.00082 & 0.046 & 0.046 & 95 &                      & -0.0031              & 0.05                 & 0.047                & 93                   &  & 0.00097              & 0.047                & 0.046                & 94                   &  & -0.00043             & 0.046                & 0.046                & 95                   \\
$\beta_{ApoE, 4}$      & 0.082                &  & 0.00021  & 0.043 & 0.041 & 93 &                      & 0.00014              & 0.043                & 0.042                & 93                   &  & 0.0024               & 0.042                & 0.041                & 95                   &  & -0.0001              & 0.044                & 0.041                & 94                   \\
$\beta_{Sex, 1}$       & -0.38                &  & -0.00096 & 0.038 & 0.037 & 94 &                      & 0.0014               & 0.039                & 0.037                & 94                   &  & -0.002               & 0.04                 & 0.037                & 93                   &  & -0.00094             & 0.038                & 0.037                & 94                   \\
$\beta_{Sex, 2}$       & 0.34                 &  & 0.0025   & 0.06  & 0.061 & 96 &                      & 0.0076               & 0.059                & 0.061                & 97                   &  & -0.0013              & 0.061                & 0.061                & 95                   &  & 0.0024               & 0.06                 & 0.061                & 96                   \\
$\beta_{Sex, 3}$       & -0.42                &  & 0.00016  & 0.047 & 0.045 & 94 &                      & -0.0015              & 0.046                & 0.045                & 94                   &  & 0.0016               & 0.045                & 0.045                & 95                   &  & 0.0002               & 0.047                & 0.045                & 94                   \\
$\beta_{Sex, 4}$       & 0.074                &  & -0.00075 & 0.037 & 0.036 & 95 &                      & -0.0013              & 0.039                & 0.036                & 93                   &  & 0.0024               & 0.042                & 0.036                & 91                   &  & -0.00074             & 0.037                & 0.036                & 95                   \\
$\beta_{Education, 1}$ & -0.1                 &  & -0.00063 & 0.018 & 0.018 & 96 &                      & -0.00066             & 0.019                & 0.018                & 96                   &  & 9e-05                & 0.019                & 0.018                & 95                   &  & -0.00062             & 0.018                & 0.018                & 96                   \\
$\beta_{Education, 2}$ & 0.11                 &  & -0.00013 & 0.03  & 0.029 & 95 &                      & 0.001                & 0.029                & 0.029                & 97                   &  & 0.00084              & 0.031                & 0.029                & 94                   &  & -0.00017             & 0.03                 & 0.029                & 95                   \\
$\beta_{Education, 3}$ & 0.096                &  & -0.00079 & 0.023 & 0.022 & 95 &                      & -0.0015              & 0.023                & 0.022                & 94                   &  & -0.00047             & 0.022                & 0.022                & 96                   &  & -0.00083             & 0.023                & 0.022                & 95                   \\
$\beta_{Education, 4}$ & 0.01                 &  & -0.00046 & 0.018 & 0.017 & 94 &                      & -0.00088             & 0.018                & 0.017                & 91                   &  & 0.00059              & 0.02                 & 0.017                & 91                   &  & -0.00049             & 0.018                & 0.018                & 95                   \\
$\gamma_{12}$          & -1.1                 &  & 0.39     & 0.21  & 0.33  & 96 &                      & 0.024                & 0.25                 & 0.25                 & 96                   &  & 0.37                 & 0.22                 & 0.33                 & 96                   &  & 0.39                 & 0.22                 & 0.33                 & 96                   \\
$\gamma_{22}$          & 0.17                 &  & -0.17    & 0.089 & 1.5   & 81 &                      & -0.5                 & 1.1                  & 2.8                  & 93                   &  & -0.17                & 0.093                & 1.5                  & 80                   &  & -0.17                & 0.09                 & 1.6                  & 81                   \\
$\gamma_{32}$          & 0.17                 &  & -0.16    & 0.073 & 1.6   & 81 &                      & -0.46                & 0.96                 & 2.4                  & 92                   &  & -0.16                & 0.073                & 1.6                  & 79                   &  & -0.17                & 0.085                & 1.5                  & 79                   \\
$\gamma_{42}$          & 0.27                 &  & -0.12    & 0.18  & 0.31  & 98 &                      & 0.046                & 0.42                 & 0.45                 & 94                   &  & -0.11                & 0.2                  & 0.31                 & 97                   &  & -0.13                & 0.29                 & 0.3                  & 87                   \\
$\gamma_{13}$          & 2.6                  &  & -0.031   & 0.11  & 0.11  & 94 &                      & 0.0055               & 0.11                 & 0.13                 & 96                   &  & -0.017               & 0.14                 & 0.11                 & 87                   &  & -0.032               & 0.11                 & 0.11                 & 93                   \\
$\gamma_{23}$          & 2.3                  &  & 0.036    & 0.22  & 0.22  & 97 &                      & 0.12                 & 0.4                  & 0.4                  & 98                   &  & 0.043                & 0.22                 & 0.22                 & 96                   &  & 0.038                & 0.25                 & 0.59                 & 97                   \\
$\gamma_{33}$          & 2.0                  &  & 0.012    & 0.17  & 0.18  & 97 &                      & 0.0061               & 0.2                  & 0.25                 & 99                   &  & 0.021                & 0.21                 & 0.19                 & 93                   &  & 0.0081               & 0.19                 & 0.19                 & 96                   \\
$\gamma_{43}$          & 1.5                  &  & -0.00077 & 0.073 & 0.077 & 95 &                      & 0.0044               & 0.074                & 0.09                 & 94                   &  & 0.0012               & 0.11                 & 0.078                & 84                   &  & -0.0014              & 0.074                & 0.077                & 95                   \\
$\sigma_1$             & 0.92                 &  & -0.0032  & 0.014 & 0.013 & 94 &                      & -0.003               & 0.014                & 0.014                & 94                   &  & -0.0027              & 0.013                & 0.013                & 94                   &  & -0.0031              & 0.014                & 0.013                & 94                   \\
$\sigma_2$             & 0.96                 &  & -0.0038  & 0.021 & 0.021 & 93 &                      & -0.0027              & 0.022                & 0.021                & 93                   &  & -0.0028              & 0.02                 & 0.021                & 95                   &  & -0.0036              & 0.021                & 0.021                & 93                   \\
$\sigma_3$             & 0.89                 &  & -0.0014  & 0.015 & 0.015 & 95 &                      & -0.0017              & 0.015                & 0.015                & 95                   &  & -0.0023              & 0.016                & 0.015                & 95                   &  & -0.0013              & 0.015                & 0.015                & 95                   \\
$\sigma_4$             & 0.63                 &  & -0.0033  & 0.014 & 0.014 & 95 &                      & -0.0027              & 0.013                & 0.014                & 97                   &  & -0.003               & 0.014                & 0.014                & 94                   &  & -0.0034              & 0.015                & 0.014                & 94                   \\
$\rho^*$                   & 0.069                &  & -0.00073 & 0.032 & 0.032 & 95 &                      & 0.0029               & 0.03                 & 0.032                & 96                   &  & 0.00041              & 0.033                & 0.032                & 95                   &  & -0.00065             & 0.032                & 0.032                & 95                   \\
$\alpha_{\text{Age}}$  & 0.72                 &  & -0.0075  & 0.053 & 0.055 & 96 &                      & 0.0069               & 0.053                & 0.06                 & 95                   &  & -0.0005              & 0.079                & 0.056                & 83                   &  & -0.0078              & 0.053                & 0.055                & 95                   \\
$\alpha_{\text{ApoE}}$ & 0.59                 &  & -0.004   & 0.088 & 0.09  & 95 &                      & 0.0035               & 0.094                & 0.1                  & 95                   &  & -0.012               & 0.16                 & 0.091                & 75                   &  & -0.0045              & 0.088                & 0.09                 & 95                   \\ \hline
\end{tabular}
}
\begin{flushleft}
    \tiny{*: $\rho$ is the correlation between the second and third biomarkers conditional on covariates.}
\end{flushleft}
\end{minipage}

\begin{minipage}{\linewidth}
\centering
\captionof{table}{Summary statistics for simulations emulating the ADNI data.}
\label{table:simulation-ADNI}
\resizebox{\linewidth}{!}{
\begin{tabular}{cccccccccccccccccccccc}
\hline\\[-0.8em]
Parameter                     & True value           &                      & Bias     & ESE   & MSE   & CP &                      & Bias     & ESE   & MSE   & CP &                      & Bias     & ESE   & MSE   & CP &                      & Bias     & ESE   & MSE   & CP \\ \hline
\multicolumn{1}{l}{}          & \multicolumn{1}{l}{} & \multicolumn{1}{l}{} & \multicolumn{4}{c}{Correct model, with penalty}          & \multicolumn{1}{l}{} & \multicolumn{4}{c}{Correct model, without penalty}          & \multicolumn{1}{l}{} & \multicolumn{4}{c}{Time correlation model, with penalty}          & \multicolumn{1}{l}{} & \multicolumn{4}{c}{Perturbed $\gamma_1$, with penalty}          \\ \cline{1-2} \cline{4-7} \cline{9-12} \cline{14-17} \cline{19-22} \\[-0.8em]
\multicolumn{1}{l}{Converged percentage}          & \multicolumn{1}{l}{} & \multicolumn{1}{l}{} & \multicolumn{4}{c}{99.0\%}          & \multicolumn{1}{l}{} & \multicolumn{4}{c}{47.9\%}          & \multicolumn{1}{l}{} & \multicolumn{4}{c}{98.8\%}          & \multicolumn{1}{l}{} & \multicolumn{4}{c}{98.1\%}\\
$\beta_{\text{Intercept}, 1}$ & 0.1                  &                      & 0.019    & 0.034 & 0.035 & 93 &                      & -0.0033  & 0.041 & 0.038 & 92 &                      & 0.017    & 0.035 & 0.035 & 92 &                      & 0.019    & 0.034 & 0.035 & 93 \\
$\beta_{\text{Intercept}, 2}$ & -0.15                &                      & -0.003   & 0.056 & 0.056 & 95 &                      & -0.027   & 0.087 & 0.077 & 96 &                      & -0.00086 & 0.052 & 0.056 & 97 &                      & -0.0028  & 0.056 & 0.056 & 94 \\
$\beta_{\text{Intercept}, 3}$ & 0.24                 &                      & -0.00093 & 0.042 & 0.041 & 94 &                      & -0.01    & 0.044 & 0.049 & 97 &                      & -0.003   & 0.041 & 0.041 & 95 &                      & -0.00062 & 0.042 & 0.041 & 95 \\
$\beta_{\text{Intercept}, 4}$ & -0.35                &                      & -0.0037  & 0.035 & 0.035 & 95 &                      & -0.00093 & 0.037 & 0.036 & 94 &                      & -0.0059  & 0.037 & 0.035 & 94 &                      & -0.0051  & 0.038 & 0.035 & 85 \\
$\beta_{\text{Age}, 1}$       & -0.13                &                      & 0.011    & 0.022 & 0.022 & 92 &                      & 0.001    & 0.023 & 0.023 & 95 &                      & 0.0098   & 0.022 & 0.022 & 92 &                      & 0.011    & 0.022 & 0.022 & 95 \\
$\beta_{\text{Age}, 2}$       & 0.11                 &                      & 0.0006   & 0.035 & 0.035 & 95 &                      & -0.0044  & 0.037 & 0.038 & 96 &                      & -0.0004  & 0.034 & 0.035 & 96 &                      & 0.00065  & 0.035 & 0.035 & 94 \\
$\beta_{\text{Age}, 3}$       & 0.18                 &                      & 0.0011   & 0.024 & 0.025 & 95 &                      & -0.0011  & 0.024 & 0.027 & 96 &                      & -0.00021 & 0.025 & 0.025 & 94 &                      & 0.0015   & 0.024 & 0.025 & 95 \\
$\beta_{\text{Age}, 4}$       & 0.15                 &                      & 0.00029  & 0.019 & 0.02  & 95 &                      & 0.00041  & 0.019 & 0.02  & 95 &                      & -0.0014  & 0.02  & 0.02  & 95 &                      & -5.9e-05 & 0.021 & 0.02  & 91 \\
$\beta_{\text{ApoE}, 1}$      & -0.013               &                      & 0.0069   & 0.04  & 0.041 & 95 &                      & -0.0048  & 0.042 & 0.043 & 95 &                      & 0.012    & 0.043 & 0.041 & 93 &                      & 0.0069   & 0.04  & 0.041 & 95 \\
$\beta_{\text{ApoE}, 2}$      & -0.044               &                      & 0.00026  & 0.062 & 0.062 & 95 &                      & -0.0035  & 0.066 & 0.064 & 95 &                      & 0.0019   & 0.06  & 0.062 & 95 &                      & 0.00044  & 0.063 & 0.063 & 95 \\
$\beta_{\text{ApoE}, 3}$      & -0.033               &                      & -0.00082 & 0.046 & 0.046 & 95 &                      & -0.0031  & 0.05  & 0.047 & 93 &                      & 0.00097  & 0.047 & 0.046 & 94 &                      & -0.00043 & 0.046 & 0.046 & 94 \\
$\beta_{\text{ApoE}, 4}$      & 0.082                &                      & 0.00021  & 0.043 & 0.041 & 93 &                      & 0.00014  & 0.043 & 0.042 & 93 &                      & 0.0024   & 0.042 & 0.041 & 95 &                      & -0.0001  & 0.044 & 0.041 & 95 \\
$\beta_{\text{Sex}, 1}$       & -0.38                &                      & -0.00096 & 0.038 & 0.037 & 94 &                      & 0.0014   & 0.039 & 0.037 & 94 &                      & -0.002   & 0.04  & 0.037 & 93 &                      & -0.00094 & 0.038 & 0.037 & 95 \\
$\beta_{\text{Sex}, 2}$       & 0.34                 &                      & 0.0025   & 0.06  & 0.061 & 96 &                      & 0.0076   & 0.059 & 0.061 & 97 &                      & -0.0013  & 0.061 & 0.061 & 95 &                      & 0.0024   & 0.06  & 0.061 & 95 \\
$\beta_{\text{Sex}, 3}$       & -0.42                &                      & 0.00016  & 0.047 & 0.045 & 94 &                      & -0.0015  & 0.046 & 0.045 & 94 &                      & 0.0016   & 0.045 & 0.045 & 95 &                      & 0.0002   & 0.047 & 0.045 & 95 \\
$\beta_{\text{Sex}, 4}$       & 0.074                &                      & -0.00075 & 0.037 & 0.036 & 95 &                      & -0.0013  & 0.039 & 0.036 & 93 &                      & 0.0024   & 0.042 & 0.036 & 91 &                      & -0.00074 & 0.037 & 0.036 & 96 \\
$\beta_{\text{Education}, 1}$ & -0.1                 &                      & -0.00063 & 0.018 & 0.018 & 96 &                      & -0.00066 & 0.019 & 0.018 & 96 &                      & 9e-05    & 0.019 & 0.018 & 95 &                      & -0.00062 & 0.018 & 0.018 & 96 \\
$\beta_{\text{Education}, 2}$ & 0.11                 &                      & -0.00013 & 0.03  & 0.029 & 95 &                      & 0.001    & 0.029 & 0.029 & 97 &                      & 0.00084  & 0.031 & 0.029 & 94 &                      & -0.00017 & 0.03  & 0.029 & 95 \\
$\beta_{\text{Education}, 3}$ & 0.096                &                      & -0.00079 & 0.023 & 0.022 & 95 &                      & -0.0015  & 0.023 & 0.022 & 94 &                      & -0.00047 & 0.022 & 0.022 & 96 &                      & -0.00083 & 0.023 & 0.022 & 96 \\
$\beta_{\text{Education}, 4}$ & 0.01                 &                      & -0.00046 & 0.018 & 0.017 & 94 &                      & -0.00088 & 0.018 & 0.017 & 91 &                      & 0.00059  & 0.02  & 0.017 & 91 &                      & -0.00049 & 0.018 & 0.018 & 96 \\
$\gamma_{12}$               & -1.1                 &                      & 0.39     & 0.21  & 0.33  & 96 &                      & 0.024    & 0.25  & 0.25  & 96 &                      & 0.37     & 0.22  & 0.33  & 96 &                      & 0.39     & 0.22  & 0.33  & 98 \\
$\gamma_{22}$               & 0.17                 &                      & -0.17    & 0.089 & 1.5   & 81 &                      & -0.5     & 1.1   & 2.8   & 93 &                      & -0.17    & 0.093 & 1.5   & 80 &                      & -0.17    & 0.09  & 1.6   & 67 \\
$\gamma_{32}$               & 0.17                 &                      & -0.16    & 0.073 & 1.6   & 81 &                      & -0.46    & 0.96  & 2.4   & 92 &                      & -0.16    & 0.073 & 1.6   & 79 &                      & -0.17    & 0.085 & 1.5   & 99 \\
$\gamma_{42}$               & 0.27                 &                      & -0.12    & 0.18  & 0.31  & 98 &                      & 0.046    & 0.42  & 0.45  & 94 &                      & -0.11    & 0.2   & 0.31  & 97 &                      & -0.13    & 0.29  & 0.3   & 85 \\
$\gamma_{13}$               & 2.6                  &                      & -0.031   & 0.11  & 0.11  & 94 &                      & 0.0055   & 0.11  & 0.13  & 96 &                      & -0.017   & 0.14  & 0.11  & 87 &                      & -0.032   & 0.11  & 0.11  & 95 \\
$\gamma_{23}$               & 2.3                  &                      & 0.036    & 0.22  & 0.22  & 97 &                      & 0.12     & 0.4   & 0.4   & 98 &                      & 0.043    & 0.22  & 0.22  & 96 &                      & 0.038    & 0.25  & 0.59  & 94 \\
$\gamma_{33}$               & 2.0                  &                      & 0.012    & 0.17  & 0.18  & 97 &                      & 0.0061   & 0.2   & 0.25  & 99 &                      & 0.021    & 0.21  & 0.19  & 93 &                      & 0.0081   & 0.19  & 0.19  & 92 \\
$\gamma_{43}$               & 1.5                  &                      & -0.00077 & 0.073 & 0.077 & 95 &                      & 0.0044   & 0.074 & 0.09  & 94 &                      & 0.0012   & 0.11  & 0.078 & 84 &                      & -0.0014  & 0.074 & 0.077 & 93 \\
$\sigma_1$                    & 0.92                 &                      & -0.0032  & 0.014 & 0.013 & 94 &                      & -0.003   & 0.014 & 0.014 & 94 &                      & -0.0027  & 0.013 & 0.013 & 94 &                      & -0.0031  & 0.014 & 0.013 & 95 \\
$\sigma_2$                    & 0.96                 &                      & -0.0038  & 0.021 & 0.021 & 93 &                      & -0.0027  & 0.022 & 0.021 & 93 &                      & -0.0028  & 0.02  & 0.021 & 95 &                      & -0.0036  & 0.021 & 0.021 & 95 \\
$\sigma_3$                    & 0.89                 &                      & -0.0014  & 0.015 & 0.015 & 95 &                      & -0.0017  & 0.015 & 0.015 & 95 &                      & -0.0023  & 0.016 & 0.015 & 95 &                      & -0.0013  & 0.015 & 0.015 & 95 \\
$\sigma_4$                    & 0.63                 &                      & -0.0033  & 0.014 & 0.014 & 95 &                      & -0.0027  & 0.013 & 0.014 & 97 &                      & -0.003   & 0.014 & 0.014 & 94 &                      & -0.0034  & 0.015 & 0.014 & 93 \\
$\rho^*$                        & 0.069                &                      & -0.00073 & 0.032 & 0.032 & 95 &                      & 0.0029   & 0.03  & 0.032 & 96 &                      & 0.00041  & 0.033 & 0.032 & 95 &                      & -0.00065 & 0.032 & 0.032 & 94 \\
$\alpha_{\text{Age}}$         & 0.72                 &                      & -0.0075  & 0.053 & 0.055 & 96 &                      & 0.0069   & 0.053 & 0.06  & 95 &                      & -0.0005  & 0.079 & 0.056 & 83 &                      & -0.0078  & 0.053 & 0.055 & 95 \\
$\alpha_{\text{ApoE}}$        & 0.59                 &                      & -0.004   & 0.088 & 0.09  & 95 &                      & 0.0035   & 0.094 & 0.1   & 95 &                      & -0.012   & 0.16  & 0.091 & 75 &                      & -0.0045  & 0.088 & 0.09  & 96 \\ \hline
\end{tabular}
}
\begin{flushleft}
    \tiny{*: $\rho$ is the correlation between the second and third biomarkers conditional on covariates.}
\end{flushleft}
\end{minipage}

%\subsection{Model specification}
%We applied the model to the marker domains and covariates specified above. Several possible covariate and covariance structures were examined, including adding/removing correlations between markers coming from the same domain, and adding/removing each of the covariates. The final structure was selected based on the Akaike Information Criterion. The tuning parameter $\lambda_n$ was selected by cross-validation. We trained the model with $\lambda_n=10^{-2},10^{-3},10^{-4}$, and chose the $\lambda_n$ which gave the lowest cross-validated error. \\
%The final covariate structure obtained based on the AIC is shown in

%{\daisy estimates, variance estimates. order }

%\subsection{Predicted Biomarker and Disease Trajectories}\label{sec:data-analysis-prediction}

\section{Discussion}\label{sec:discussion}
In this paper, we propose a non-linear mixed effect model to formulate Jack and his colleagues' biomarker cascade hypothesis for AD. Our proposed methods take into consideration the unique data structures often seen in observational longitudinal studies of AD, such as the BIOCARD and ADNI. Through simulation studies, we investigate the finite-sample behavior of the proposed methods under various settings with challenges that are often found in the real data. We then use the proposed estimation methods to characterize the biomarker cascades using the BIOCARD and ADNI data. The analysis results suggest that CSF biomarkers tend to worsen first, followed by MRI and cognitive biomarkers, which are patterns consistent with current evidence in the literature and existing hypotheses. We also proposed prediction approaches to monitor disease progression and biomarker deterioration, and illustrate the procedures using one participant's data from the BIOCARD study.

Our work provides a means to study AD biomarker changes in relation to the underlying disease progression that are not directly observable, and thus has important implications for clinical practice and future research related to diagnosing and predicting AD, especially preclinical AD. However, there are a few limitations to consider. 

Firstly, we adopt a parametric model to potentially enhance interpretability of results and conclusive information on biomarker ordering given the limited available data. Biases can occur if the parametric assumption is misspecified, and the results need to be interpreted with this limitation. In comparison, semi-parametric or non-parametric modeling approaches may provide a more flexible way to study the complex relationships between biomarkers and disease progression when a larger cohort and more longitudinal biomarker measurements are available over a long follow-up period. However, even with richer data, it might still be necessary to impose certain restrictions based on current understandings of AD for improving estimation efficiency and obtaining analytic results that might contribute to AD research. 

Secondly, we analyzed the BIOCARD and ADNI data separately, recognizing the potential batch effect and the differences in the cognitive normal cohorts enrolled in these two studies. Specifically, the ADNI cohort demonstrated greater homogeneity in terms of cognitive function compared to the BIOCARD cohort. This difference could have complex implications on the distributions of biomarkers and covariates. Combining the two study datasets without appropriate harmonization methods may produce misleading results. Meanwhile, separately analyzing the two datasets allowed us to independently study and validate biomarker cascading results. However, such analyses resulted in a smaller sample size and potential loss of efficiency. In addition, the covariates and biomarkers were standardized within each study, and the latent disease scales were defined differently for the two studies. This affected the generalizability of our results across the two studies and potentially to a patient not recruited in BIOCARD or ADNI. Future research could explore methods to harmonize and combine both datasets to improve the precision of estimates, and potentially increase the generalizability of the results. 

Thirdly, we proposed some prediction and monitoring methodologies based on our model to enhance our understanding of the multi-faceted biomarker profile tied to AD progression and to facilitate the identification of aggressive disease advancement and biomarker deterioration. However, it is noteworthy that these predictions are not linked or calibrated to existing diagnostic criteria. As a result, interpreting disease scores and biomarker abnormality levels in clinical practice can be challenging. Future endeavors would be helpful to establish a link between our model results and the current clinical practice, enhancing the practical application of our research in disease management and treatment decision-making.

\section*{Acknowledgements}
This work is partially supported by the National Institutes of Health’s National Institute on Aging grant R01AG068002. We greatly appreciate Dr. Shannon L. Cole for his invaluable contributions in revising and editing this manuscript, which greatly improved its quality.
We are grateful for the support of the entire BIOCARD study team at Johns Hopkins University, the ADNI study team, and to the dedicated participants who continue to participate in these studies. The BIOCARD study is supported in part by grants from the National Institutes of Health: U01-AG033655, P50-AG005146 and P41-RR015241. The BIOCARD Study consists of 7 Cores with the following members: (1) the Administrative Core (Marilyn Albert, Rostislav Brichko), (2) the Clinical Core (Marilyn Albert, Anja Soldan, Corinne Pettigrew, Rebecca Gottesman, Ned Sacktor, Scott Turner, Leonie Farrington, Maura Grega, Gay Rudow, Scott Rudow), (3) the Imaging Core (Michael Miller, Susumu Mori, Tilak Ratnanather, Anthony Kolasny, Kenichi Oishi, Laurent Younes), (4) the Biospecimen Core (Abhay Moghekar, Jacqueline Darrow, Richard O’Brien), (5) the Informatics Core (Roberta Scherer, David Shade, Ann Ervin, Jennifer Jones, Hamadou Coulibaly, April Patterson), the (6) Biostatistics Core (Mei-Cheng Wang, Daisy Zhu, Jiangxia Wang), and (7) the Neuropathology Core (Juan Troncoso, Olga Pletnikova, Gay Rudow, Karen Fisher).

The ADNI data collection and sharing for this project was funded by the Alzheimer's Disease Neuroimaging Initiative (ADNI) (National Institutes of Health Grant U01-AG024904) and DOD ADNI (Department of Defense award number W81XWH-12-2-0012). ADNI is funded by the National Institute on Aging, the National Institute of Biomedical Imaging and Bioengineering, and through generous contributions from the following: AbbVie, Alzheimer's Association; Alzheimer's Drug Discovery Foundation; Araclon Biotech; BioClinica, Inc.; Biogen; Bristol-Myers Squibb Company; CereSpir, Inc.; Cogstate; Eisai Inc.; Elan Pharmaceuticals, Inc.; Eli Lilly and Company; EuroImmun; F. Hoffmann-La Roche Ltd and its affiliated company Genentech, Inc.; Fujirebio; GE Healthcare; IXICO Ltd.; Janssen Alzheimer Immunotherapy Research \& Development, LLC.; Johnson \& Johnson Pharmaceutical Research \& Development LLC.; Lumosity; Lundbeck; Merck \& Co., Inc.; Meso Scale Diagnostics, LLC.; NeuroRx Research; Neurotrack Technologies; Novartis Pharmaceuticals Corporation; Pfizer Inc.; Piramal Imaging; Servier; Takeda Pharmaceutical Company; and Transition Therapeutics. The Canadian Institutes of Health Research is providing funds to support ADNI clinical sites in Canada. Private sector contributions are facilitated by the Foundation for the National Institutes of Health (www.fnih.org). The grantee organization is the Northern California Institute for Research and Education, and the study is coordinated by the Alzheimer's Therapeutic Research Institute at the University of Southern California. ADNI data are disseminated by the Laboratory for Neuro Imaging at the University of Southern California.

\bibliography{AD_LDA}
\bibliographystyle{apalike}

\end{document}